\documentclass[sigconf, screen]{acmart}
\usepackage{graphicx}
\usepackage{bbm, dsfont}
\usepackage{subcaption}
\usepackage{algorithm}
\usepackage{algpseudocode}
\usepackage{multirow}
\usepackage{threeparttable}

\AtBeginDocument{%
  \providecommand\BibTeX{{%
    \normalfont B\kern-0.5em{\scshape i\kern-0.25em b}\kern-0.8em\TeX}}}

\setcopyright{acmlicensed}
\copyrightyear{2024}
\acmYear{2024}
\acmDOI{XXXXXXX.XXXXXXX}
\begin{document}

\title{Joint Contrastive Learning with Feature Alignment for Cross-Corpus EEG-based Emotion Recognition}

% \affiliation{%

\author{Qile Liu}
\email{liuqile2022@email.szu.edu.cn}
\affiliation{
  \institution{Shenzhen University}
  \city{Shenzhen}
  \country{China}
}

\author{Zhihao Zhou}
\email{2310247057@email.szu.edu.cn}
\affiliation{
  \institution{Shenzhen University}
  \city{Shenzhen}
  \country{China}
}

\author{Jiyuan Wang}
\email{2310247016@email.szu.edu.cn}
\affiliation{
  \institution{Shenzhen University}
  \city{Shenzhen}
  \country{China}
}

\author{Zhen Liang}
\authornote{Corresponding author.}
\email{janezliang@szu.edu.cn}
\affiliation{
  \institution{Shenzhen University}
  \city{Shenzhen}
  \country{China}
}

% \author{Anonymous Authors}
%% You do not have to enter your paper ID


%% The abstract is a short summary of the work to be presented in the article.
\begin{abstract}
The integration of human emotions into multimedia applications shows great potential for enriching user experiences and enhancing engagement across various digital platforms. Unlike traditional methods such as questionnaires, facial expressions, and voice analysis, brain signals offer a more direct and objective understanding of emotional states. However, in the field of electroencephalography (EEG)-based emotion recognition, previous studies have primarily concentrated on training and testing EEG models within a single dataset, overlooking the variability across different datasets. This oversight leads to significant performance degradation when applying EEG models to cross-corpus scenarios. In this study, we propose a novel \textbf{J}oint \textbf{C}ontrastive learning framework with \textbf{F}eature \textbf{A}lignment (\textbf{JCFA}) to address cross-corpus EEG-based emotion recognition. The JCFA model operates in two main stages. In the pre-training stage, a joint domain contrastive learning strategy is introduced to characterize generalizable time-frequency representations of EEG signals, without the use of labeled data. It extracts robust time-based and frequency-based embeddings for each EEG sample, and then aligns them within a shared latent time-frequency space. In the fine-tuning stage, JCFA is refined in conjunction with downstream tasks, where the structural connections among brain electrodes are considered. The model capability could be further enhanced for the application in emotion detection and interpretation. Extensive experimental results on two well-recognized emotional datasets show that the proposed JCFA model achieves state-of-the-art (SOTA) performance, outperforming the second-best method by an average accuracy increase of 4.09\% in cross-corpus EEG-based emotion recognition tasks.
\end{abstract}




\maketitle

\section{Introduction}
Objective assessment of an individual’s emotional states is of great importance in the field of human-computer interaction, disease diagnosis and rehabilitation \cite{cowie2001emotion, fragopanagos2005emotion, carpenter2018cognitive, zotev2020emotion}. Existing research mainly uses two types of data for emotion recognition: behavioral signals and physiological signals. Behavioral signals, including speech \cite{gu2018deep}, gestures \cite{noroozi2018survey}, and facial expressions \cite{zeng2018facial}, are low-cost and easily accessible but can be intentionally altered. On the other hand, physiological signals, such as electrocardiography (ECG), electromyography (EMG), and electroencephalography (EEG), offer a more reliable measure of emotions as they are less susceptible to manipulation. EEG, in particular, stands out for its ability to provide direct, objective insights into emotional states by capturing brain activity from different locations on the scalp. Therefore, researchers from diverse fields have increasingly focused on EEG-based emotion recognition in recent years \cite{liu2017real, li2021multi, gong2023astdf, li2022eeg}.

Currently, a variety of methods have been developed for EEG-based emotion recognition. For example, Song \textit{et al.} \cite{song2018eeg} proposed a dynamical graph convolutional neural network (DGCNN) to dynamically learn the intrinsic relationship among brain electrodes. To capture both local and global relations among different EEG channels, Zhong \textit{et al.} \cite{zhong2020eeg} proposed a regularized graph neural network (RGNN). Concerning domain adaptation techniques, Li \textit{et al.} \cite{li2019domain} introduced a joint distribution network (JDA) that adapts the joint distributions by simultaneously adapting marginal distributions and conditional distributions. Incorporating self-supervised learning, Shen \textit{et al.} \cite{shen2022contrastive} proposed a contrastive learning method for inter-subject alignment (CLISA), and achieved state-of-the-art (SOTA) performance in cross-subject EEG-based emotion recognition tasks. More related literature review is available in Section \ref{sec:RelatedWork}. However, two critical challenges remain unaddressed in current methods. \textbf{(1) Experimental protocol.} Existing approaches mainly consider within-subject and cross-subject experimental protocols within one single dataset, neglecting to account for variations between different datasets. This oversight can significantly reduce the effectiveness of established methods when applied to cross-corpus scenarios \cite{rayatdoost2018cross}. \textbf{(2) Data availability.} Zhou \textit{et al.} \cite{zhou2023eeg} introduced an EEG-based emotion style transfer network (E$^2$STN) to integrate content information from the source domain with style information from the target domain, achieving promising performance in cross-corpus scenarios. However, it requires access to all labeled source data and unlabeled target data beforehand. Considering the difficulties in collecting EEG signals and the expert knowledge and time required to label them, obtaining labeled data in advance for model training is impractical in real-world applications.

To tackle the aforementioned two critical challenges, this paper proposes a novel \textbf{J}oint \textbf{C}ontrastive learning framework with \textbf{F}eature \textbf{A}lignment (\textbf{JCFA}) for cross-corpus EEG-based emotion recognition. The proposed JCFA model is a self-supervised learning approach that includes two main stages. In the pre-training stage, a joint contrastive learning strategy is introduced to identify aligned time- and frequency-based embeddings of EEG signals that remain consistent across diverse environmental conditions present in different datasets. In the fine-tuning stage, a further enhancement on model capability to the downstream tasks is developed by incorporating spatial features of brain electrodes using a graph convolutional network. In summary, the main contributions of our work are outlined as follows: 
\begin{itemize}
\item We propose a novel JCFA framework designed to tackle two main critical challenges (experimental protocol and data availability) encountered in the field of cross-corpus EEG-based emotion recognition.
\item We introduce a joint contrastive learning strategy to align time- and frequency-based embeddings and produce generalizable EEG feature representations, all without the need of labeled data.
\item We conduct extensive experiments on two well-recognized datasets comparing with 10 existing methods, showing JCFA achieves SOTA performance with an average accuracy increase of 4.09\% over the second-best method.
\end{itemize}

\section{Related Work}
In this section, we review the related work in terms of EEG-based emotion recognition, cross-corpus EEG emotion recognition, and contrastive learning.
\label{sec:RelatedWork}
\subsection{EEG-based Emotion Recognition}
Existing methods for EEG-based emotion recognition can be categorised into two groups. \textbf{(1) Machine learning-based methods} use manually extracted EEG features, such as Power Spectral Density (PSD) \cite{goldfischer1965autocorrelation} and Differential Entropy (DE) \cite{duan2013differential}, for emotion recognition. For example, Alsolamy \textit{et al.} \cite{alsolamy2016emotion} employed PSD features as input to a support vector machine (SVM) \cite{suykens1999least} in emotion recognition. However, the performance of traditional machine learning-based methods tends to be relatively poor and unstable. Therefore, researchers have turned to developing various \textbf{(2) deep learning-based models} in recent years. For example, Song \textit{et al.} \cite{song2021graph} proposed a graph-embedded convolutional neural network (GECNN) to extract discriminative spatial features. Additionally, they introduced a variational instance-adaptive graph method (V-IAG) \cite{song2021variational}, which captures the individual dependencies among different electrodes and estimates the underlying uncertain information. To further reduce the impact of individual differences, transfer learning strategy is suggested to incorporate. For instance, Li \textit{et al.} \cite{li2018novel} proposed a bi-hemispheres domain adversarial neural network (BiDANN) to address domain shifts between different subjects. Considering the local and global feature distributions, Li \textit{et al.} \cite{li2021novel} proposed a transferable attention neural network (TANN) for learning discriminative information using attention mechanism. Further, Zhou \textit{et al.} \cite{zhou2023pr} introduced a novel transfer learning framework with prototypical representation based pairwise learning (PR-PL) to address individual differences and noisy labeling in EEG emotional data, achieving SOTA performance. However, the above methods would fail in adapting to cross-corpus scenarios, due to the substantial differences in feature representation among different datasets.

\subsection{Cross-Corpus EEG Emotion Recognition}
Considering practical requirements, researchers have recently begun to explore EEG-based emotion recognition in cross-corpus scenarios. For example, Rayatdoost \textit{et al.} \cite{rayatdoost2018cross} developed a deep convolutional network and evaluated the performance of existing methods in cross-corpus scenarios. Their study demonstrated a significant decline in model performance when these methods were evaluated across different corpora. Lan \textit{et al.} \cite{lan2018domain} also performed a comparative analysis to evaluate the existing domain adaptation methods in cross-corpus scenarios. Their findings indicated that although domain adaptation techniques improved upon baseline models, the overall model performance remained less than ideal, with classification accuracies ranging from 30\% to 50\%. To facilitate the application of domain adaptation techniques in cross-corpus scenarios, Zhou \textit{et al.} \cite{zhou2023eeg} introduced an EEG-based emotion style transfer network (E$^2$STN), which integrates content information from the source domain with style information from the target domain, and achieved SOTA performance in multiple cross-corpus scenarios. However, the existing methods necessitate access to all labeled source data and unlabeled target data for model training in advance, making it impractical to apply them in real applications.

\begin{figure*}
    \begin{center}
        \includegraphics[width=0.85\textwidth]{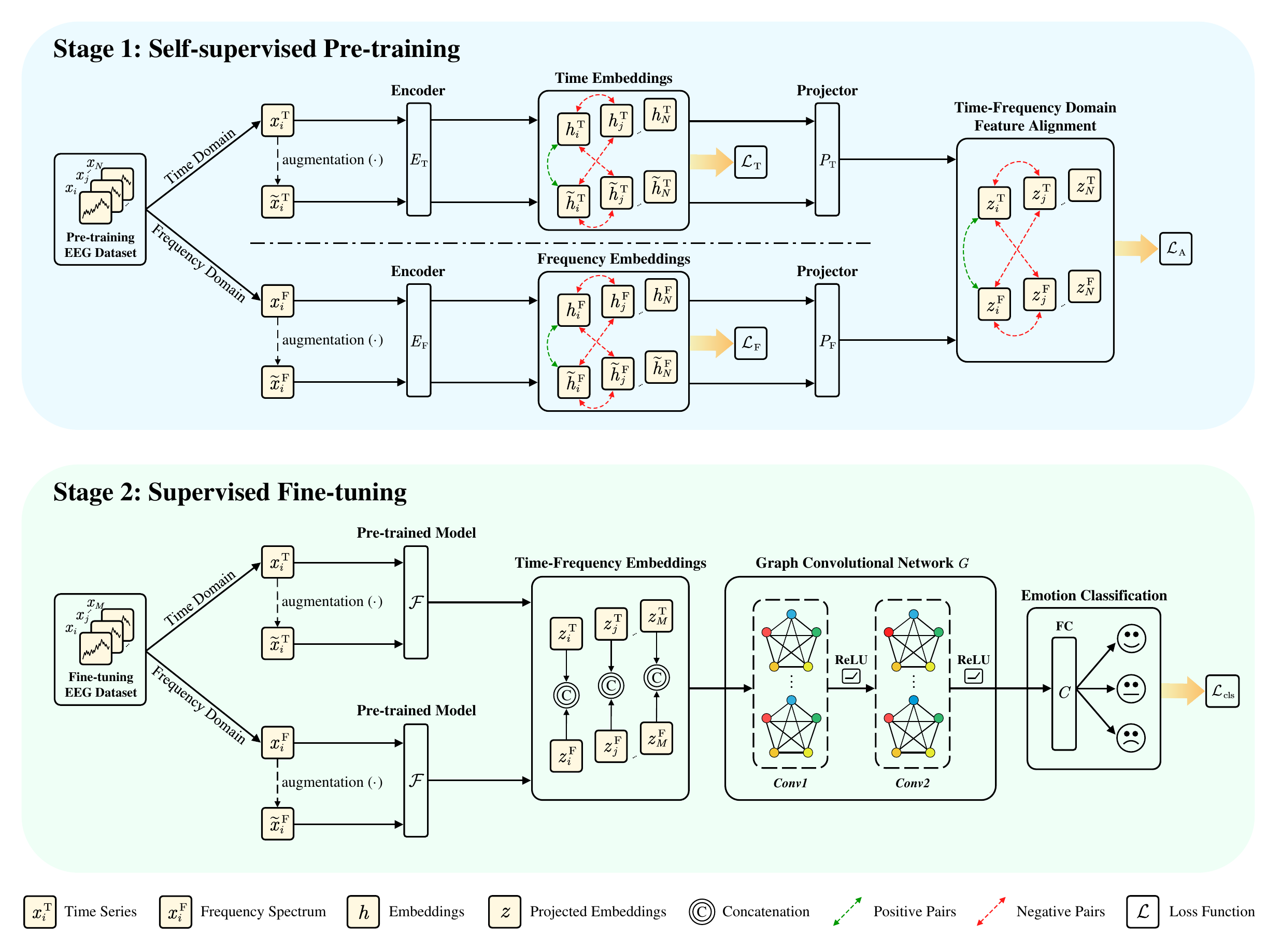}
    \end{center}
    \caption{The overall architecture of the proposed JCFA model for cross-corpus EEG-based emotion recognition. JCFA consists of two stages: (1) joint contrastive learning-based self-supervised pre-training stage, and (2) graph convolutional network-based supervised fine-tuning stage.}
    \label{fig:Overall Framework}
    \Description{This figure illustrates the overall architecture of the proposed JCFA for cross-corpus EEG-based emotion recognition.}
\end{figure*}

\subsection{Contrastive Learning}
Contrastive learning is a popular type of self-supervised learning, which has achieved superior performance in feature representation without requiring labeled data. Recently, researchers have started applying contrastive learning to physiological signals. For example, Wickstr{\o}m \textit{et al.} \cite{wickstrom2022mixing} introduced a new mixing-up augmentation method for ECG signals, extending the concept of Mixup \cite{zhang2017mixup} from computer vision to time series analysis. For EEG analysis, inspired by SimCLR \cite{chen2020simple}, Mohsenvand \textit{et al.} \cite{mohsenvand2020contrastive} proposed a contrastive learning model for three tasks: sleep stage classification, clinical abnormal detection, and emotion recognition. Moreover, Shen \textit{et al.} \cite{shen2022contrastive} proposed a contrastive learning method for inter-subject alignment (CLISA) and significantly improved the performance of cross-subject EEG-based emotion recognition. To date, the applicability of self-supervised contrastive learning in cross-corpus scenarios has not been explored. Therefore, in this paper, we introduce a joint contrastive learning framework to efficiently extract time- and frequency-based embedding information that remains reliable across EEG signals collected from different datasets.

\section{Problem Formulation}
\label{sec:Problem Formulation}
For an unlabeled pre-training dataset $\mathcal{D}^{\mathrm{pret}} = \{x_i^{\mathrm{pret}} \mid i = 1,\ldots,N \}$, which consists of $N$ samples with $V^{\mathrm {pret}}$ channels and $T^{\mathrm {pret}}$ timestamps. For a labeled fine-tuning dataset $\mathcal{D}^{\mathrm{tune}} = \{ (x_i^{\mathrm{tune}}, y_i) \mid i = 1,\ldots,M \}$, which contains $M$ samples with $V^{\mathrm {tune}}$ channels and $T^{\mathrm {tune}}$ timestamps. Each sample $x_i^{\mathrm{tune}}$ is paired with a label $y_i \in \{ 1,\ldots,C \}$, where $C$ is the number of emotion categories. The objective is to pre-train a model $\mathcal{F}$, utilizing $\mathcal{D}^{\mathrm{pret}}$ through a self-supervised learning strategy. The pre-training process enables $\mathcal{F}$ to derive generalizable representations of EEG signals, without the need of labeled information. Then, we refined the pre-trained model $\mathcal{F}$ in the supervised fine-tuning stage, using a small amount of the labeled data from $\mathcal{D}^{\mathrm{tune}}$. The fine-tuned model $\mathcal{F}$ can be regarded as adept at well performing cross-corpus EEG-based emotion recognition.

\textbf{The central idea} of the JCFA model is inspired by a fundamental assumption in signal processing theory \cite{flandrin1998time, papandreou2018applications}, which suggests the existence of a latent time-frequency space. In this space, the time-based embeddings and frequency-based embeddings learned from the same time series should be closer to each other compared to embeddings from different time series. To satisfy this assumption, a joint contrastive learning strategy across time, frequency, and time-frequency domains is proposed. Here, the extracted time- and frequency-based embeddings of the same sample are aligned within the time-frequency domain. The resulting time- and frequency-based embeddings are considered generalizable representations of EEG signals, which are robust across various datasets.

\section{Methodology}
In this section, we propose a JCFA framework for cross-corpus EEG-based emotion recognition. The overall architecture of JCFA is shown in Fig. \ref{fig:Overall Framework}. It consists of two stages: (1) joint contrastive learning-based self-supervised pre-training stage and (2) graph convolutional network-based supervised fine-tuning stage. 

\subsection{Self-supervised Pre-training}
In this part, we explain the design of our pre-training stage, which uses the self-supervised joint contrastive learning strategy. $x_i^{\rm T}$ and $x_i^{\rm F}$ denote 
the input time domain and frequency domain raw data of an EEG sample $\boldsymbol{\mathit{x_i}}$. For simplicity, we use univariate (single-channel) EEG signals as an example below to elucidate the pre-training process. Note that our approach can be straightforwardly applied to multivariate (multi-channel) EEG signals. The detailed self-supervised pre-training process is illustrated in Algorithm \ref{alg:pre-training}.

\subsubsection{Time Domain Contrastive Learning.} 
For an input univariate EEG sample $\boldsymbol{\mathit{x_i}} \in \mathcal{D}^{\mathrm{pret}}$, we first conduct weak augmentation using $x_i^{\rm T}$ to generate the corresponding augmented sample $\widetilde{x}_i^{\rm T}$. Here, we perform jittering operation by adding Gaussian noise as weak augmentation. Then, we feed $x_i^{\rm T}$ and $\widetilde{x}_i^{\rm T}$ into a time-based contrastive encoder $E_{\rm T}$ that maps samples to time-based embeddings, denoted as $h_i^{\rm T} = E_{\rm T}(x_i^{\rm T})$ and $\widetilde{h}_i^{\rm T} = E_{\rm T}(\widetilde{x}_i^{\rm T})$. Since $\widetilde{x}_i^{\rm T}$ is generated by $x_i^{\rm T}$ through data augmentation, after the time encoder $E_{\rm T}$, we assume that $h_i^{\rm T}$ (the embedding of $x_i^{\rm T}$) should be close to $\widetilde{h}_i^{\rm T}$ (the embedding of $\widetilde{x}_i^{\rm T}$), but far away from $h_j^{\rm T}$ and $\widetilde{h}_j^{\rm T}$ (the embeddings of another sample $x_j^{\rm T}$ and its augmentation $\widetilde{x}_j^{\rm T}$) \cite{chen2020simple, kiyasseh2021clocs, yue2022ts2vec}. Thus, we define $(x_i^{\rm T}, \widetilde{x}_i^{\rm T})$ as positive pair, and $(x_i^{\rm T}, x_j^{\rm T})$ and $(x_i^{\rm T}, \widetilde{x}_j^{\rm T})$ as negative pairs \cite{chen2020simple}. To maximize the similarity between positive pairs and minimize the similarity between negative pairs, we adopt the NT-Xent (the normalized temperature-scaled cross entropy loss) \cite{chen2020simple, tang2020exploring} to measure the distance between sample pairs. In particular, for a positive pair $(x_i^{\rm T}, \widetilde{x}_i^{\rm T})$, the time-based contrastive loss is defined as:
\begin{equation}
\label{eq:Time-based Contrastive Loss}
    \mathcal{L}_{{\rm T},i}
                ={\rm -log}\frac{{\rm exp}({\rm sim}(h_i^{\rm T}, \widetilde{h}_i^{\rm T})/\tau)}
                {\sum_{\boldsymbol{\mathit{x_j}} \in \mathcal{D}^{\mathrm{pret}}} \mathds{1}_{i \neq j} {\rm exp}({\rm sim}(h_i^{\rm T}, E_{\rm T}(\boldsymbol{\mathit{x_j}}))/\tau)},
\end{equation}
where $\rm sim(\cdot,\cdot)$ refers to the cosine similarity. $\mathds{1}_{i \neq j}$ is an indicator function that equals to 1 when $i \neq j$ and 0 otherwise. $\tau$ is a temperature parameter. The $\boldsymbol{\mathit{x_j}} \in \mathcal{D}^{\mathrm{pret}}$ represents another sample $x_j^{\rm T}$ and its augmentation $\widetilde{x}_j^{\rm T}$ that are different from $\boldsymbol{\mathit{x_i}}$. Note that $\mathcal{L}_{{\rm T},i}$ is computed across all positive pairs, both $(x_i^{\rm T}, \widetilde{x}_i^{\rm T})$ and $(\widetilde{x}_i^{\rm T}, x_i^{\rm T})$. During the training process, the final time-based contrastive loss $\mathcal{L}_{\rm T}$ is computed as the average of $\mathcal{L}_{{\rm T},i}$ within a mini-batch.

\subsubsection{Frequency Domain Contrastive Learning.}
\label{subsubsection:frequncy contrastive learning} 
To exploit the frequency information in EEG signals, we design a frequency-based contrastive encoder $E_{\rm F}$ to capture robust frequency-based embeddings. First, we generate the frequency spectrum $x_i^{\rm F}$ from $x_i^{\rm T}$ by Fourier Transformation \cite{nussbaumer1982fast}. Then, we perform data augmentation in frequency domain by perturbing the frequency spectrum. The existing research has shown that applying small perturbations to the frequency spectrum can easily lead to significant changes in the original time series \cite{flandrin1998time}. Thus, unlike the substantial perturbation by adding Gaussian noise directly to frequency spectrum in \cite{liu2021contrastive}, we perturb the frequency spectrum $x_i^{\rm F}$ weakly by removing and adding a small portion of frequency components. Specifically, we generate a probability matrix $U$ drawn from a uniform distribution $U(0,1)$, which matches the dimensionality of $x_i^{\rm F}$. When removing frequency components, we zero out the amplitudes in $x_i^{\rm F}$ at locations where the values in $U$ are smaller than $\mu$. When adding frequency components, we replace the amplitudes in $x_i^{\rm F}$ with $\epsilon \cdot A_m$ at locations where the values in $U$ are greater than (1 - $\mu$). Here, $\mu$ is a probability threshold that controls the range of spectrum perturbations, $\epsilon$ is a scaling factor ($\mu$ and $\epsilon$ are set to 0.05 and 0.1 empirically), and $A_m$ is the maximum amplitude in the frequency spectrum.

Through removing and adding a small portion of frequency components, we generate the corresponding augmented sample $\widetilde{x}_i^{\rm F}$ from the original $x_i^{\rm F}$. Then, both $x_i^{\rm F}$ and $\widetilde{x}_i^{\rm F}$ are inputted into a frequency-based contrastive encoder $E_{\rm F}$, which transforms inputs into the frequency-based embeddings, represented as $h_i^{\rm F} = E_{\rm F}(x_i^{\rm F})$ and $\widetilde{h}_i^{\rm F} = E_{\rm F}(\widetilde{x}_i^{\rm F})$. We assume that the frequency encoder $E_{\rm F}$ is capable of generating similar frequency embeddings for $x_i^{\rm F}$ and $\widetilde{x}_i^{\rm F}$. Accordingly, $(x_i^{\rm F}, \widetilde{x}_i^{\rm F})$ is defined as positive pair, and $(x_i^{\rm F}, x_j^{\rm F})$ and $(x_i^{\rm F}, \widetilde{x}_j^{\rm F})$ are considered as negative pairs. Similar to the time domain contrastive learning, for a positive pair $(x_i^{\rm F}, \widetilde{x}_i^{\rm F})$, the frequency-based contrastive loss is defined as:
\begin{equation}
\label{eq:Frequency-based Contrastive Loss}
    \mathcal{L}_{{\rm F},i}
                ={\rm -log}\frac{{\rm exp}({\rm sim}(h_i^{\rm F}, \widetilde{h}_i^{\rm F})/\tau)}
                {\sum_{\boldsymbol{\mathit{x_j}} \in \mathcal{D}^{\mathrm{pret}}} \mathds{1}_{i \neq j} {\rm exp}({\rm sim}(h_i^{\rm F}, E_{\rm F}(\boldsymbol{\mathit{x_j}}))/\tau)}.
\end{equation}

Similar to $\mathcal{L}_{{\rm T},i}$, we calculate $\mathcal{L}_{{\rm F},i}$ among all positive pairs, including $(x_i^{\rm F}, \widetilde{x}_i^{\rm F})$ and $(\widetilde{x}_i^{\rm F}, x_i^{\rm F})$. In the training process, the final frequency-based contrastive loss $\mathcal{L}_{\rm F}$ is computed as the average of $\mathcal{L}_{{\rm F},i}$ within a mini-batch. The $\mathcal{L}_{\rm F}$ urges the frequency encoder $E_{\rm F}$ to produce embeddings that are robust to spectrum perturbations.

\subsubsection{Time-Frequency Domain Contrastive Learning.} 
To enable the model to capture generalizable time-frequency representations of EEG signals, we conduct time-frequency domain contrastive learning and introduce an alignment loss $\mathcal{L}_{\rm A}$ to synchronize the time- and frequency-based embeddings. First, we map $h_i^{\rm T}$ from time domain and $h_i^{\rm F}$ from frequency domain into a shared latent time-frequency space using two cross-space projectors $P_{\rm T}$ and $P_{\rm F}$, respectively. In particular, we have two time-based and frequency-based embeddings for each sample $\boldsymbol{\mathit{x_i}}$ through $P_{\rm T}$ and $P_{\rm F}$, which are denoted as $z_i^{\rm T} = P_{\rm T}(h_i^{\rm T})$ and $z_i^{\rm F} = P_{\rm F}(h_i^{\rm F})$. Then, we assume that the time-based embedding and frequency-based embedding learned from the same sample should be closer to each other in the latent space than embeddings of different samples. Therefore, we define the time-based embedding and frequency-based embedding from the same sample as positive pair. The time- and frequency-based embeddings from different samples are considered as negative pairs. In other words, $(z_i^{\rm T}, z_i^{\rm F})$ is positive pair, while $(z_i^{\rm T}, z_j^{\rm T})$ and $(z_i^{\rm T}, z_j^{\rm F})$ are negative pairs. To fulfill the key idea of JCFA, for a positive pair $(z_i^{\rm T}, z_i^{\rm F})$, we define the alignment loss $\mathcal{L}_{{\rm A},i}$ as:
\begin{equation}
\label{eq:Alignment Loss}             
    \mathcal{L}_{{\rm A},i}
                ={\rm -log}\frac{{\rm exp}({\rm sim}(z_i^{\rm T},z_i^{\rm F})/\tau)}
                {\sum_{\boldsymbol{\mathit{z_j}} \in \mathcal{Z}^{\mathrm{pret}}} \mathds{1}_{i \neq j} {\rm exp}({\rm sim}(z_i^{\rm T}, \boldsymbol{\mathit{z_j}})/\tau)},
\end{equation}
where $\mathcal{Z}^{\mathrm{pret}}$ represents the shared latent time-frequency space. $\boldsymbol{\mathit{z_j}} \in \mathcal{Z}^{\mathrm{pret}}$ denotes the projected time- and frequency-based embeddings ($z_j^{\rm T}$ and $z_j^{\rm F}$) from another sample $\boldsymbol{\mathit{x_j}}$. Here, $\mathcal{L}_{{\rm A},i}$ is computed across all positive pairs, including $(z_i^{\rm T}, z_i^{\rm F})$ and $(z_i^{\rm F}, z_i^{\rm T})$. We calculate the final alignment loss $\mathcal{L}_{\rm A}$ as the average of $\mathcal{L}_{{\rm A},i}$ within a mini-batch. Under the constraint of $\mathcal{L}_{\rm A}$, the model is encouraged to align the time- and frequency-based embeddings for each EEG sample within the shared latent time-frequency space.
\begin{algorithm}[h]
    \caption{The algorithmic flow of the pre-training stage.}
    \label{alg:pre-training}
    \begin{algorithmic}[1]
        \Require
    \renewcommand{\algorithmicrequire}{\textbf{}}
    \Require - Unlabeled pre-training EEG emotion dataset $\mathcal{D}^{\mathrm{pret}} = \{ x_i^{\mathrm{pret}} \mid i = 1,\ldots,N \}$. The number of pre-training $epochs$. 
    \Ensure
    \State Random initialization of model parameters;
    \Statex \textcolor{gray}{\# Data Augmentation}
    \For {$i=1$ to $N$}
        \State Generate $x_i^{\rm F}$ by Fourier Transformation using $x_i^{\rm T}$ in $\mathcal{D}^{\mathrm{pret}}$;
        \State Perform time augmentation for $x_i^{\rm T}$ to generate $\widetilde{x}_i^{\rm T}$; 
        \State Perform frequency augmentation for $x_i^{\rm F}$ to generate $\widetilde{x}_i^{\rm F}$;
    \EndFor
    \State \textcolor{gray}{\# Model Pre-training Process}
    \For {1 to $epochs$} 
        \Statex\hspace*{1em} \textcolor{gray}{\# \# All operations are performed within a mini-batch}
        \Statex\hspace*{1em} \textcolor{gray}{\# Time Domain Contrastive Learning}
        \State Generate $h_i^{\rm T}$ and $\widetilde{h}_i^{\rm T}$ through inputting $x_i^{\rm T}$ and $\widetilde{x}_i^{\rm T}$ into $E_{\rm T}$;
        \State Calculate the time-based contrastive loss $\mathcal{L}_{\rm T}$ in \textbf{Eq. \ref{eq:Time-based Contrastive Loss}};
        \Statex\hspace*{1em} \textcolor{gray}{\# Frequency Domain Contrastive Learning}
        \State Generate $h_i^{\rm F}$ and $\widetilde{h}_i^{\rm F}$ through inputting $x_i^{\rm F}$ and $\widetilde{x}_i^{\rm F}$ into $E_{\rm F}$;
        \State Calculate the frequency-based contrastive loss $\mathcal{L}_{\rm F}$ in \textbf{Eq. \ref{eq:Frequency-based Contrastive Loss}};
        \Statex\hspace*{1em} \textcolor{gray}{\# Time-Frequency Domain Contrastive Learning}
        \State Generate $z_i^{\rm T}$ through inputting $h_i^{\rm T}$ into $P_{\rm T}$;
        \State Generate $z_i^{\rm F}$ through inputting $h_i^{\rm F}$ into $P_{\rm F}$;
        \State Calculate the alignment loss $\mathcal{L}_{\rm A}$ in \textbf{Eq. \ref{eq:Alignment Loss}};
        \Statex\hspace*{1em} \textcolor{gray}{\# Overall Pre-training Loss}
        \State Calculate the overall pre-training loss $\mathcal{L}_{\rm pret}$ in \textbf{Eq. \ref{eq:Loss of model pre-training}};
        \State Update model parameters by gradient back-propagation;
    \EndFor
    \end{algorithmic}
\end{algorithm}

During the pre-training stage, the model $\mathcal{F}$ is trained by jointly optimizing the time-based contrastive loss $\mathcal{L}_{\rm T}$, the frequency-based contrastive loss $\mathcal{L}_{\rm F}$, and the alignment loss $\mathcal{L}_{\rm A}$. The overall loss function for model pre-training is defined as:
\begin{equation}
\label{eq:Loss of model pre-training}
    \mathcal{L}_{{\rm pret}}
            =\alpha (\mathcal{L}_{\rm T} + \mathcal{L}_{\rm F}) + \beta \mathcal{L}_{\rm A},
\end{equation}
where $\alpha$ and $\beta$ are two pre-defined constants that control the contribution of the contrastive and alignment losses.

\subsection{Supervised Fine-tuning}
In the fine-tuning stage, we define a graph convolutional network $G$ to enhance the capture of spatial features from time- and frequency-based embeddings derived from brain electrodes. Specifically, for an input multi-channel EEG sample $\boldsymbol{\mathit{x_i}} \in \mathcal{D}^{\mathrm{tune}}$, we input each channel of $\boldsymbol{\mathit{x_i}}$ into the pre-trained model $\mathcal{F}$ separately to produce the corresponding time-based embedding $z_{i,j}^{\rm T}$ and frequency-based embedding $z_{i,j}^{\rm F}$. Here, $j$ is the channel order and $j \in \{1,\ldots,V^{\rm {tune}}\}$. By aggregating $z_{i,j}^{\rm T}$ as well as $z_{i,j}^{\rm F}$ from all channels, we can obtain the final time embedding $z_i^{\rm T}$ and frequency embedding $z_i^{\rm F}$ of sample $\boldsymbol{\mathit{x_i}}$. Then, we concatenate $z_i^{\rm T}$ and $z_i^{\rm F}$ into a joint time-frequency embedding, denoted as $Z_i = [z_i^{\rm T}; z_i^{\rm F}] \in \mathbb{R}^{V^{\mathrm {tune}} \times F}$, where $F$ is the feature dimension. By utilizing $Z_i$, the pre-trained model $\mathcal{F}$ is further fine-tuned in a supervised manner using a small subset of labeled data from $\mathcal{D}^{\mathrm{tune}}$. This fine-tuning process involves the integration of the graph convolutional network $G$ and an emotion classifier $C$. We provide the detailed algorithmic flow of the supervised fine-tuning process in Algorithm \ref{alg:fine-tuning}. The impact of selecting a small subset for fine-tuning is carefully examined in Section \ref{sec:impact of fine-tuning size}.

\subsubsection{Graph Convolutional Network.} 
To further capture the spatial features of EEG signals, we integrate a graph convolutional network $G$ in the fine-tuning stage. Inspired by previous work \cite{song2018eeg, jin2023pgcn}, we design a cosine similarity-based distance function to better describe the connectivity between different nodes in $Z_i$ rather than using simple numbers 0 and 1. Then, we construct an initial adjacency matrix $\boldsymbol{\mathit{A}} \in \mathbb{R}^{V^{\mathrm {tune}} \times V^{\mathrm {tune}}}$, each element of which is expressed as:
\begin{equation}
    \label{eq:Adjacency Matrix Based on Cosine Similarity}
        A_{mn}=
            \begin{cases}
                {\rm exp}({\rm sim}(m, n) - 1) & \text{if } \delta \leq A_{mn} \leq 1 \\
                \delta & \text{if } 0 \leq A_{mn} < \delta
            \end{cases}.
\end{equation}

We clip $A_{mn}$ with $\delta$ to ensure that weak connectivity still exists between nodes $m$ and $n$ with low similarity. Based on the obtained adjacency matrix $\boldsymbol{\mathit{A}}$, we use the graph convolution based on spectral graph theory to capture spatial features of brain electrodes. Specifically, we adopt the $K$-order Chebyshev graph convolution \cite{defferrard2016convolutional} for graph $G$ considering the computational complexity. Here, we define $G$ with only two layers to avoid over-smoothing.

\subsubsection{Emotion Classification.} 
The final objective of JCFA is to achieve accurate emotion classification in cross-corpus scenarios. We reshape the outputs of $G$ into one-dimensional vectors, and input them into an emotion classifier $C$ consisting of a 3-layer fully connected network for final emotion recognition. The classifier $C$ is trained by minimizing the cross entropy loss between the ground truth and predicted labels:
\begin{equation}
\label{eq:Loss of Classifier}
    \mathcal{L}_{{\rm cls}} = 
        \frac{1}{M} \sum_{m=1}^{M} \sum_{c=1}^C -y_{m,c} {\rm log}(p_{m,c}) - (1-y_{m,c} {\rm log}(p_{m,c})),
\end{equation}
where $p_{m,c}$ is the probability that the $m$-th sample belongs to $c$-th class and $y_{m,c} \in \{0,1\}$. $M$ is the total number of samples used in model fine-tuning.

During the fine-tuning process, we train $G$ and $C$ with the optimization of $\mathcal{L}_{\rm cls}$, while jointly optimizing $\mathcal{L}_{\rm T}$, $\mathcal{L}_{\rm F}$ and $\mathcal{L}_{\rm A}$ to fine-tune the pre-trained model $\mathcal{F}$. In summary, the overall fine-tuning loss is defined as:
\begin{equation}
\label{eq:Loss of model fine-tuning}
    \mathcal{L}_{{\rm tune}}
            =\alpha (\mathcal{L}_{\rm T} + \mathcal{L}_{\rm F}) + \beta \mathcal{L}_{\rm A} + 
             \gamma \mathcal{L}_{\rm cls},
\end{equation}
where $\alpha$, $\beta$ and $\gamma$ are pre-defined constants controlling the weights of each loss. Note that $\mathcal{L}_{\rm T}$, $\mathcal{L}_{\rm F}$ and $\mathcal{L}_{\rm A}$ are computed as the average of the corresponding loss over all channels.
\begin{algorithm}
    \caption{The algorithmic flow of the fine-tuning stage.}
    \label{alg:fine-tuning}
    \begin{algorithmic}[1]
        \Require
    \renewcommand{\algorithmicrequire}{\textbf{}}
    \Require - Labeled fine-tuning EEG emotion dataset $\mathcal{D}^{\mathrm{tune}} = \{ (x_i^{\mathrm{tune}}, y_i) \mid i = 1,\ldots,M \}$. The number of fine-tuning $epochs$. 
    \Ensure
    \State Initializing model parameters using the pre-trained model $\mathcal{F}$;
    \State Perform data augmentation for $\boldsymbol{\mathit{x_i}} \in \mathcal{D}^{\mathrm{tune}}$;
    \State \textcolor{gray}{\# Model Fine-tuning Process}
    \For {1 to $epochs$} 
        \Statex\hspace*{1em} \textcolor{gray}{\# \# All operations are performed within a mini-batch}
        \State \parbox[t]{\dimexpr\linewidth-\algorithmicindent} {Generate the corresponding $z_{i,j}^{\rm T}$ and $z_{i,j}^{\rm F}$ through inputting each channel of $\boldsymbol{\mathit{x_i}}$ into $\mathcal{F}$ separately;}
        \State \parbox[t]{\dimexpr\linewidth-\algorithmicindent} {Calculate the average of $\mathcal{L}_{\rm T}$, $\mathcal{L}_{\rm F}$ and $\mathcal{L}_{\rm A}$ over all channels using \textbf{Eq. \ref{eq:Time-based Contrastive Loss}}, \textbf{Eq. \ref{eq:Frequency-based Contrastive Loss}}, and \textbf{Eq. \ref{eq:Alignment Loss}};}
        \State \parbox[t]{\dimexpr\linewidth-\algorithmicindent} {Aggregate $z_{i,j}^{\rm T}$ and $z_{i,j}^{\rm F}$ from all channels into the final time embedding $z_i^{\rm T}$ and frequency embedding $z_i^{\rm F}$, respectively;}
        \State Concatenate $z_i^{\rm T}$ and $z_i^{\rm F}$ into the feature matrix $Z_i$;
        \Statex\hspace*{1em} \textcolor{gray}{\# $K$-order Chebyshev Graph Convolution}
        \State Calculate the cosine similarity between each channel of $Z_i$;
        \State Construct the adjacency matrix $\boldsymbol{\mathit{A}}$ using \textbf{Eq. \ref{eq:Adjacency Matrix Based on Cosine Similarity}};
        \State \parbox[t]{\dimexpr\linewidth-\algorithmicindent}{Generate spatial features using $K$-order Chebyshev graph convolution based on the adjacency matrix $\boldsymbol{\mathit{A}}$;}
        \Statex\hspace*{1em} \textcolor{gray}{\# Emotion Classification}
        \State \parbox[t]{\dimexpr\linewidth-\algorithmicindent} {Flatten the node features into one-dimensional vectors and feed them into the classifier to predict the categories;}
        \State Calculate the classification loss $\mathcal{L}_{{\rm cls}}$ in \textbf{Eq. \ref{eq:Loss of Classifier}};
        \Statex\hspace*{1em} \textcolor{gray}{\# Overall Fine-tuning Loss}
        \State Calculate the overall fine-tuning loss $\mathcal{L}_{\rm tune}$ in \textbf{Eq. \ref{eq:Loss of model fine-tuning}};
        \State Update model parameters by gradient back-propagation;
    \EndFor
    \end{algorithmic}
\end{algorithm}

\section{Experiments}
\label{sec:Experiments}
\subsection{Datasets}
We conduct extensive experiments on two well-known datasets, SEED \cite{zheng2015investigating} and SEED-IV \cite{zheng2018emotionmeter}, to evaluate the performance of our proposed JCFA model in cross-corpus scenarios. In the experiments, we only consider the case where the pre-training and fine-tuning datasets have the same emotion categories (i.e., negative, neutral and positive emotions), and we use the preprocessed 1-s EEG signals for both datasets. More details about datasets are presented in \ref{appendix:Datasets}.

\subsection{Implementation Details}
We use two 2-layer transformer encoders as backbones for the time-based contrastive encoder $E_{\rm T}$ and the frequency-based contrastive encoder $E_{\rm F}$, without parameters sharing. The two cross-space projectors $P_{\rm T}$ and $P_{\rm F}$ are composed of two 2-layer fully connected networks, with no sharing parameters. In the pre-training stage, we set $\tau$ to a small value of 0.05 to increase the penalization of hard negative samples. We set $\alpha$ and $\beta$ in Eq. \ref{eq:Loss of model pre-training} to 0.2 and 1, respectively. In the fine-tuning stage, we set $K$ to a small value of 3 in $G$ to avoid over-smoothing. Then, we set $\tau$ to 0.5, $\alpha$ and $\beta$ to 0.1, and $\gamma$ to 1. The model pre-training and fine-tuning processes are conducted for 1000 and 20 epochs, respectively. We use a batch size of 256 for pre-training and 128 for fine-tuning. The proposed JCFA model is implemented using Python 3.9 and trained with PyTorch 1.13 on an NVIDIA GeForce RTX 3090 GPU. More implementation details are in \ref{appendix:Implementation details}.

\subsection{Baseline Models and Experimental Settings}
We conduct a comprehensive comparison of the proposed JCFA model with 10 existing methods, including 8 conventional methods: SVM \cite{suykens1999least}, DANN \cite{ganin2016domain}, A-LSTM \cite{song2019mped}, V-IAG \cite{song2021variational}, GECNN \cite{song2021graph}, BiDANN \cite{li2018novel}, TANN \cite{li2021novel} and E$^2$STN \cite{zhou2023eeg}, and 2 classical contrastive learning models: SimCLR \cite{chen2020simple, tang2020exploring} and Mixup \cite{zhang2017mixup, wickstrom2022mixing}. More details about baseline models are in \ref{appendix:Baselines}. 

To ensure a fair comparison, we use the default hyper-parameters reported in the original paper, and follow the same experimental settings (cross-corpus subject-independent protocol) for all the above methods. Accordingly, we can obtain two kinds of experimental results: training with SEED-IV and testing on SEED (SEED-IV$^3$ $\rightarrow$ SEED$^3$), and training with SEED and testing on SEED-IV (SEED$^3$ $\rightarrow$ SEED-IV$^3$). We use accuracy (ACC) as the evaluation metric for model performance. All methods are evaluated using the average and standard deviation of the results for all subjects in the test set. Further information about experimental settings is available in \ref{appendix:Baselines}.

\begin{table}
\caption{3-category classification accuracies (mean ± std \%) for cross-corpus EEG-based emotion recognition on the SEED and SEED-IV datasets.}
    \centering
    \resizebox{\columnwidth}{!}{
        \begin{threeparttable}
        \begin{tabular}{c|c|c}
        \hline
        \hline
            \multirow{2}{*}{\rule{0pt}{10pt} Methods} & \multicolumn{2}{c}{ACC ± STD (\%)} \\
            \cline{2-3} & \rule{0pt}{10pt} SEED-IV$^3$ $\rightarrow$ SEED$^3$ & \rule{0pt}{10pt} SEED$^3$ $\rightarrow$ SEED-IV$^3$ \\
            \hline
            SVM \cite{suykens1999least}* & 41.77 ± 11.13 & 29.57 ± 00.73 \\
            DANN \cite{ganin2016domain}* & 49.95 ± 09.27 & 44.53 ± 03.60 \\ 
            A-LSTM \cite{song2019mped} & 46.47 ± 08.30 & 58.19 ± 13.73 \\ 
            V-IAG \cite{song2021variational} & 52.84 ± 07.71 & 59.87 ± 11.16 \\
            GECNN \cite{song2021graph} & 58.02 ± 07.03 & 57.25 ± 07.53 \\
            BiDANN \cite{li2018novel} & 49.24 ± 10.49 & 60.46 ± 11.17 \\
            TANN \cite{li2021novel} & 58.41 ± 07.16 & 60.75 ± 10.61 \\
            E$^2$STN \cite{zhou2023eeg} & \underline{60.51 ± 05.41} & \underline{61.24 ± 15.14} \\
            SimCLR \cite{chen2020simple, tang2020exploring}* & 47.27 ± 08.44 & 46.89 ± 13.41 \\
            Mixup \cite{zhang2017mixup, wickstrom2022mixing}* & 56.86 ± 16.83 & 55.70 ± 16.28 \\
            \textbf{JCFA (Ours)} & 
                    \textbf{67.53 ± 12.36} \textbf{\color{red}{\footnotesize(+07.02)}} & \textbf{62.40 ± 07.54} \textbf{\color{red}{\footnotesize(+01.16)}} \\
        \hline
        \hline
        \end{tabular}
        \begin{tablenotes}
            \footnotesize 
            \item * indicates the results are obtained by our own implementation. A$^3$ $\rightarrow$ B$^3$ denotes that A is the source training set and B is the target test set, where superscript $^3$ indicates 3-category classification. \textbf{Best results} across each dataset are in bold, while the \underline{second-best results} are underlined.
        \end{tablenotes}
        \end{threeparttable}
        }
\label{tab:Model performance comparison}
\end{table}

\subsection{Results Analysis and Comparison}
Table \ref{tab:Model performance comparison} shows the experimental comparison results of the proposed JCFA model with existing methods for cross-corpus EEG-based emotion recognition on the SEED and SEED-IV datasets. Here, for the SEED-IV$^3$ $\rightarrow$ SEED$^3$ experiment, we use the first 9 trials of each subject in the SEED dataset for model fine-tuning, and the last 6 trials for testing. For the SEED$^3$ $\rightarrow$ SEED-IV$^3$ experiment, we take the first 12 trials of each subject in the SEED-IV dataset for fine-tuning and the last 6 trials for testing. The experimental results demonstrate that the proposed JCFA model achieves SOTA performance on the SEED and SEED-IV datasets compared to 10 existing methods. Specifically, JCFA achieves a classification accuracy of 67.53\% with a standard deviation of 12.36\% for the SEED-IV$^3$ $\rightarrow$ SEED$^3$ experiment, outperforming the second-best method E$^2$STN by 7.02\% of accuracy. Meanwhile, JCFA achieves a recognition accuracy of 62.40\% with a standard deviation of 7.54\% for the SEED$^3$ $\rightarrow$ SEED-IV$^3$ experiment, surpassing E$^2$STN by 1.16\% in accuracy. Besides, JCFA outperforms the two classical contrastive learning models, SimCLR and Mixup, which indicates that the frequency information as well as the time-frequency domain contrastive learning with alignment loss play a significant role in improving the accuracy of emotion recognition.

\begin{figure}
    \begin{center}
        \includegraphics[width=0.42\textwidth]{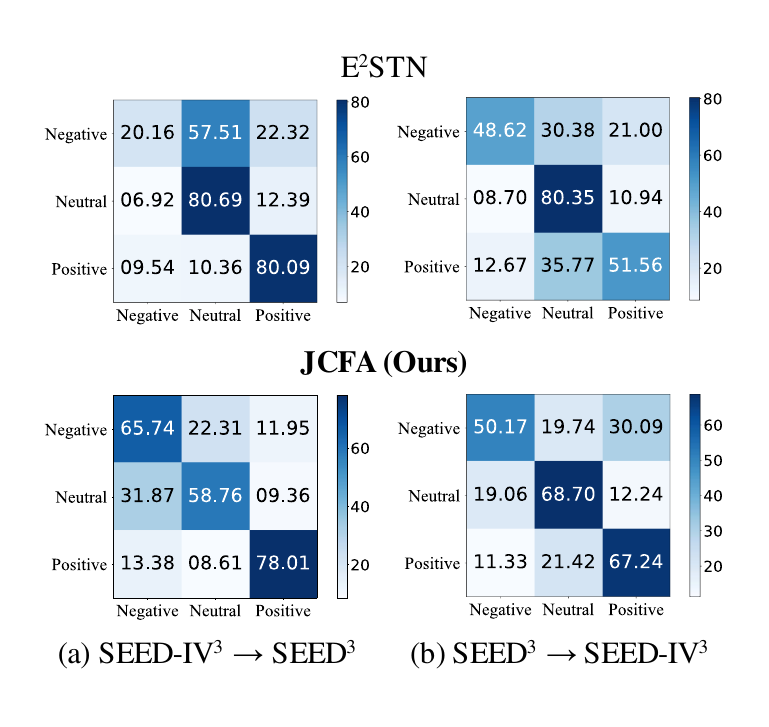}
    \end{center}
    \caption{Confusion matrices of E$^2$STN (the second-best method) and the JCFA model for cross-corpus EEG-based emotion recognition on the SEED and SEED-IV datasets.}
    \label{fig:Comparison_matrix}
    \Description{This figure illustrates the classification confusion matrices of the proposed JCFA for cross-corpus EEG-based emotion recognition.}
\end{figure}

Figure \ref{fig:Comparison_matrix} compares the classification confusion matrices of E$^2$STN (the second-best method) and the proposed JCFA model in the SEED-IV$^3$ $\rightarrow$ SEED$^3$ and SEED$^3$ $\rightarrow$ SEED-IV$^3$ experiments. The experimental results show that our model exhibits greater stability compared to E$^2$STN, achieving high classification accuracies for all three emotion categories. Notably, JCFA achieves higher recognition accuracies for the most challenging-to-recognize negative emotions in both experimental settings.

\begin{table*}
\caption{Classification performance (mean ± std \%) of ablation study for cross-corpus EEG-based emotion recognition in the SEED-IV$^3$ $\rightarrow$ SEED$^3$ experiment.}
    \centering
    \renewcommand{\arraystretch}{}
        \begin{tabular}{c|c|c|c|c|c|c}
        \hline
        \hline
            Methods & Accuracy & Precision & Recall & F1 Score & AUROC & AUPRC \\
        \hline
            w/o $\mathcal{L}_{\rm F}$, $\mathcal{L}_{\rm A}$ and $G$ & 48.68 ± 11.26 & 49.62 ± 11.99 & 48.63 ± 11.16 & 47.97 ± 11.06 & 66.24 ± 11.25 & 52.27 ± 14.05 \\
            w/o $\mathcal{L}_{\rm T}$, $\mathcal{L}_{\rm A}$ and $G$ & 63.09 ± 13.97 & 63.53 ± 14.06 & 62.87 ± 13.99 & 62.51 ± 14.27 & 79.18 ± 11.13 & 68.11 ± 14.41 \\ 
            w/o $\mathcal{L}_{\rm A}$ and $G$ & 63.38 ± 14.22 & 63.77 ± 14.71 & 63.18 ± 14.22 & 62.29 ± 15.07 & 78.19 ± 11.10 & 67.57 ± 14.78 \\
            w/o $G$ & 66.10 ± 12.74 & 66.59 ± 12.98 & 65.87 ± 12.75 & 64.94 ± 13.83 & 80.85 ± 10.11 & 70.80 ± 13.57 \\
            \textbf{Full Model (JCFA)} & \textbf{67.53 ± 12.36} & \textbf{68.12 ± 12.84} & \textbf{67.33 ± 12.44} & \textbf{66.57 ± 12.25} & \textbf{82.63 ± 10.06} & \textbf{72.46 ± 13.49} \\
        \hline
        \hline
        \end{tabular}
\label{tab:Ablation study on SEED}
\end{table*}

\begin{table*}
\caption{Classification performance (mean ± std \%) of ablation study for cross-corpus EEG-based emotion recognition in the SEED$^3$ $\rightarrow$ SEED-IV$^3$ experiment.}
    \centering
    \renewcommand{\arraystretch}{}
        \begin{tabular}{c|c|c|c|c|c|c}
        \hline
        \hline
            Methods & Accuracy & Precision & Recall & F1 Score & AUROC & AUPRC \\
        \hline
            w/o $\mathcal{L}_{\rm F}$, $\mathcal{L}_{\rm A}$ and $G$ & 46.73 ± 05.07 & 44.85 ± 04.34 & 44.30 ± 05.64 & 41.95 ± 05.24 & 61.59 ± 06.67 & 47.84 ± 06.26 \\
            w/o $\mathcal{L}_{\rm T}$, $\mathcal{L}_{\rm A}$ and $G$ & 55.89 ± 08.36 & 55.92 ± 08.38 & 55.57 ± 08.26 & 54.71 ± 08.43 & 74.79 ± 07.60 & 62.58 ± 09.30 \\ 
            w/o $\mathcal{L}_{\rm A}$ and $G$ & 57.86 ± 07.99 & 57.18 ± 07.59 & 56.96 ± 07.76 & 56.06 ± 07.58 & 74.98 ± 07.41 & 62.71 ± 08.79 \\
            w/o $G$ & 60.14 ± 06.52 & 60.12 ± 06.54 & 59.29 ± 06.48 & 58.53 ± 06.56 & \textbf{78.53 ± 05.98} & 66.32 ± 07.53 \\
            \textbf{Full Model (JCFA)} & \textbf{62.40 ± 07.54} & \textbf{61.80 ± 08.78} & \textbf{60.93 ± 08.21} & \textbf{60.09 ± 09.04} & 78.35 ± 07.74 & \textbf{66.77 ± 09.66} \\
        \hline
        \hline
        \end{tabular}
\label{tab:Ablation study on SEED-IV}
\end{table*}

\begin{table*}
\caption{Classification performance (mean ± std \%) for cross-corpus EEG-based emotion recognition in the SEED-IV$^3$ $\rightarrow$ SEED$^3$ experiment under different fine-tuning sizes.}
    \centering
    \renewcommand{\arraystretch}{}
        \begin{tabular}{c|c|c|c|c|c|c}
        \hline
        \hline
            Fine-tuning Trials & Accuracy & Precision & Recall & F1 Score & AUROC & AUPRC \\
        \hline
            3 & 60.82 ± 09.33 & 61.65 ± 09.00 & 60.64 ± 09.35 & 60.10 ± 09.46 & 78.02 ± 08.89 & 66.70 ± 11.17 \\
            6 & 63.77 ± 12.12 & 63.82 ± 12.53 & 63.40 ± 12.03 & 63.01 ± 12.26 & 79.95 ± 10.25 & 68.48 ± 13.69 \\ 
            \textbf{9} & \textbf{67.53 ± 12.36} & \textbf{68.12 ± 12.84} & \textbf{67.33 ± 12.44} & \textbf{66.57 ± 12.25} & \textbf{82.63 ± 10.06} & \textbf{72.46 ± 13.49} \\
            12 & 66.22 ± 13.72 & 67.25 ± 12.74 & 65.97 ± 13.67 & 65.07 ± 14.63 & 81.37 ± 10.58 & 71.08 ± 13.89 \\
        \hline
        \hline
        \end{tabular}
\label{tab:impact of fine-tuning size on SEED}
\end{table*}

\begin{table*}
\caption{Classification performance (mean ± std \%) for cross-corpus EEG-based emotion recognition in the SEED$^3$ $\rightarrow$ SEED-IV$^3$ experiment under different fine-tuning sizes.}
    \centering
    \renewcommand{\arraystretch}{}
        \begin{tabular}{c|c|c|c|c|c|c}
        \hline
        \hline
            Fine-tuning Trials & Accuracy & Precision & Recall & F1 Score & AUROC & AUPRC \\
        \hline
            3 & 51.60 ± 05.74 & 51.38 ± 05.85 & 51.06 ± 05.71 & 50.36 ± 05.87 & 69.20 ± 05.73 &55.82 ± 06.19 \\
            6 & 56.39 ± 05.84 & 56.20 ± 05.60 & 55.46 ± 06.22 & 54.87 ± 06.20 & 73.79 ± 06.13 & 61.03 ± 07.07 \\ 
            9 & 60.04 ± 06.55 & 59.35 ± 06.79 & 59.21 ± 06.65 & 58.43 ± 06.74 & 76.20 ± 05.84 & 64.27 ± 07.13 \\
            \textbf{12} & \textbf{62.40 ± 07.54} & 61.80 ± 08.78 & 60.93 ± 07.21 & 60.09 ± 09.04 & 78.35 ± 07.74 & \textbf{66.77 ± 09.66} \\
            15 & 61.46 ± 06.77 & \textbf{61.90 ± 06.65} &\textbf{61.35 ± 06.48} & \textbf{60.45 ± 07.05} & \textbf{78.72 ± 06.31} & 66.75 ± 07.32 \\
        \hline
        \hline
        \end{tabular}
\label{tab:impact of fine-tuning size on SEED-IV}
\end{table*}

\subsection{Ablation Study}
We conduct a comprehensive ablation study in the SEED-IV$^3$ $\rightarrow$ SEED$^3$ and SEED$^3$ $\rightarrow$ SEED-IV$^3$ experiments. To fully assess the validity of each module in our proposed JCFA model, we calculate six evaluation metrics: Accuracy, Precision, Recall, F1 Score, AUROC, and AUPRC. Specifically, we design four different models below. \textbf{(1) w/o $\mathcal{L}_{\rm F}$, $\mathcal{L}_{\rm A}$ and $G$:} we train $E_{\rm T}$ and $P_{\rm T}$ by optimizing $\mathcal{L}_{\rm T}$ to evaluate the model performance with only time domain contrastive learning. \textbf{(2) w/o $\mathcal{L}_{\rm T}$, $\mathcal{L}_{\rm A}$ and $G$:} we train $E_{\rm F}$ and $P_{\rm F}$ by optimizing $\mathcal{L}_{\rm F}$ to evaluate the model performance with only frequency domain contrastive learning. \textbf{(3) w/o $\mathcal{L}_{\rm A}$ and $G$:} we train $E_{\rm T}$, $P_{\rm T}$, $E_{\rm F}$ and $P_{\rm F}$ by optimizing $\mathcal{L}_{\rm T}$ and $\mathcal{L}_{\rm F}$ to validate the model performance without time-frequency domain contrastive learning and graph convolutional network. \textbf{(4) w/o $G$:} we train $E_{\rm T}$, $P_{\rm T}$, $E_{\rm F}$ and $P_{\rm F}$ without $G$ by optimizing $\mathcal{L}_{\rm T}$, $\mathcal{L}_{\rm F}$ and $\mathcal{L}_{\rm A}$ to validate the model performance without graph convolutional network.

Table \ref{tab:Ablation study on SEED} and Table \ref{tab:Ablation study on SEED-IV} show the experimental results of ablation study, which indicates that by effectively fusing all the modules, the proposed JCFA model achieves the best performance in both SEED-IV$^3$ $\rightarrow$ SEED$^3$ and SEED$^3$ $\rightarrow$ SEED-IV$^3$ experiments. Specifically, \textbf{(1) w/o $\mathcal{L}_{\rm F}$, $\mathcal{L}_{\rm A}$ and $G$:} if we only perform time domain contrastive learning, the model performance is worst with classification accuracies of 48.68\% and 46.73\% on the SEED and SEED-IV datasets, respectively. \textbf{(2) w/o $\mathcal{L}_{\rm T}$, $\mathcal{L}_{\rm A}$ and $G$:} if we only perform frequency domain contrastive learning, the model performance is better than that with only time domain contrastive learning, with the corresponding classification accuracies of 63.09\% and 55.89\%. It indicates the importance of frequency information in EEG-based emotion recognition tasks. \textbf{(3) w/o $\mathcal{L}_{\rm A}$ and $G$:} incorporating both time domain and frequency domain contrastive learning leads to improved model performance, achieving classification accuracies of 63.38\% and 57.86\% on the SEED and SEED-IV datasets. \textbf{(4) w/o $G$:} removing graph convolutional network negatively impacts the model due to the lack of spatial feature analysis. Furthermore, comparing the performance of models with and without time-frequency alignment reveals that aligning time- and frequency-based embeddings enhances generalizable representation extraction of EEG signals. This enhancement is particularly beneficial in addressing the challenges presented in cross-corpus scenarios.

\section{Discussions}
\subsection{Impact of Fine-tuning Size}
\label{sec:impact of fine-tuning size}
We investigate the impact of varying the number of trials used in the fine-tuning stage on model performance. Here, we define the number of fine-tuning trials as $N_{\rm T}$. Then, the samples from the remaining (15 - $N_{\rm T}$) trials in the SEED dataset and (18 - $N_{\rm T}$) trials in the SEED-IV dataset are allocated for testing. The experimental results in Table \ref{tab:impact of fine-tuning size on SEED} and Table \ref{tab:impact of fine-tuning size on SEED-IV} show that increasing the number of fine-tuning trials can effectively improve the model performance. In particular, the JCFA model achieves the highest classification accuracy on the SEED and SEED-IV datasets when $N_{\rm T}$ is equal to 9 and 12, respectively. On the other hand, further increasing the fine-tuning size may result in negative growth when $N_{\rm T}$ reaches a certain level. Excessive fine-tuning samples can introduce additional noise, which may interfere with the model learning process.

\subsection{Visualization of Time-Frequency Domain Contrastive Learning}
\begin{figure}
    \begin{center}
        \subfloat{
            \begin{minipage}{0.22\textwidth}
            \centering
                \includegraphics[width=1\textwidth]{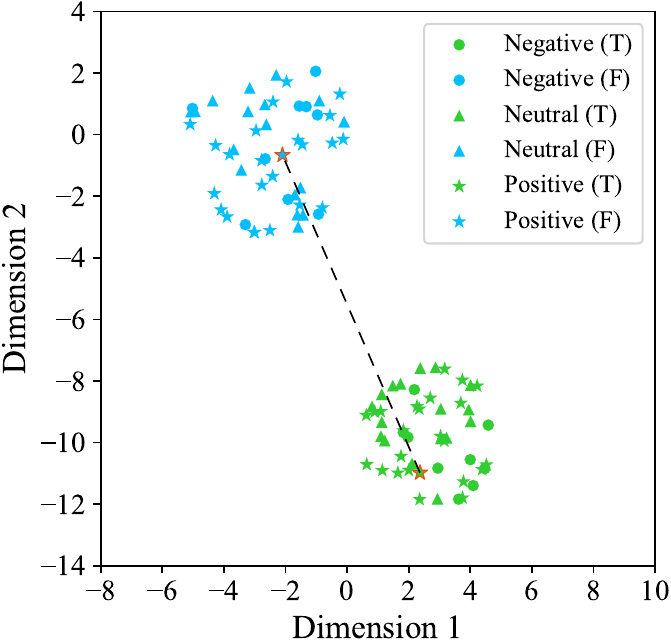}
                (a) W/o alignment loss $\mathcal{L}_{\rm A}$
            \centering
            \end{minipage}
        }
        \hspace{1em}
        \subfloat{
            \begin{minipage}[c]{0.213\textwidth}
            \centering
                \includegraphics[width=1\textwidth]{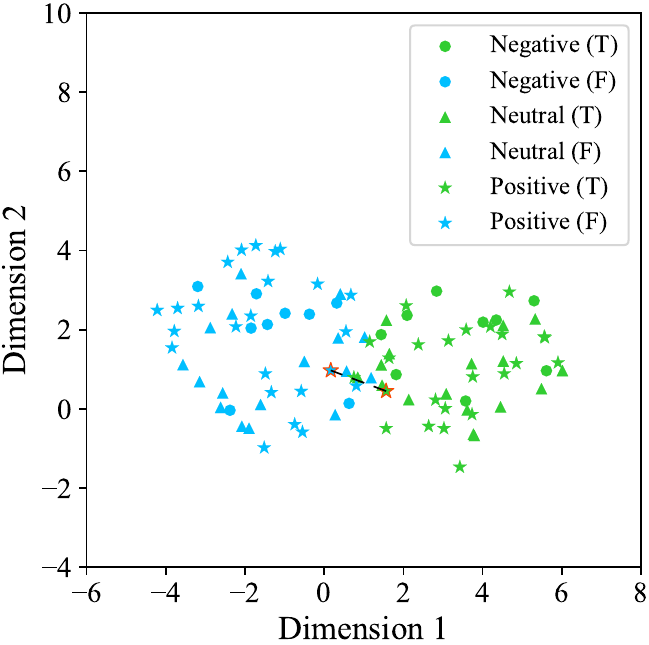}
                (b) With alignment loss $\mathcal{L}_{\rm A}$
            \centering
            \end{minipage}
        }
    \end{center}
    \caption{t-SNE visualization of embeddings in the latent time-frequency space. Circles, triangles and pentagrams denote negative, neutral and positive emotions. Green and blue represent the time embeddings and frequency embeddings. The sample marked with a red line is the same sample in two subplots. The dashed line indicates the distance between time- and frequency-based embeddings of the same sample.}
    \label{fig:feature visualization}
    \Description{This figure illustrates the visualization results of time embedding and frequency embedding using t-SNE.}
\end{figure}

To illustrate the effectiveness of the time-frequency domain contrastive learning with alignment loss, we randomly select 50 samples from the SEED dataset, and employ t-SNE \cite{van2008visualizing} to visualize the time-based embeddings (colored as green) and frequency-based embeddings (colored as blue) in the shared latent time-frequency space, as shown in Fig. \ref{fig:feature visualization}. Without the alignment loss $\mathcal{L}_{\rm A}$, the time embedding $z_i^{\rm T}$ and frequency embedding $z_i^{\rm F}$ from the same sample (marked with a red line) exhibit considerable separation. With the inclusion of the alignment loss $\mathcal{L}_{\rm A}$, the spatial gap between the time embedding $z_i^{\rm T}$ and frequency embedding $z_i^{\rm F}$ originating from the same sample is diminished (highlighted with a dashed line), signifying a notable alignment between the time- and frequency-based embeddings. The observations demonstrate the efficacy of the time-frequency domain contrastive learning with alignment loss $\mathcal{L}_{\rm A}$ in bringing closer together the time embedding $z_i^{\rm T}$ and the frequency embedding $z_i^{\rm F}$ of the identical sample.

\begin{figure}
    \begin{center}
        \includegraphics[width=0.3\textwidth]{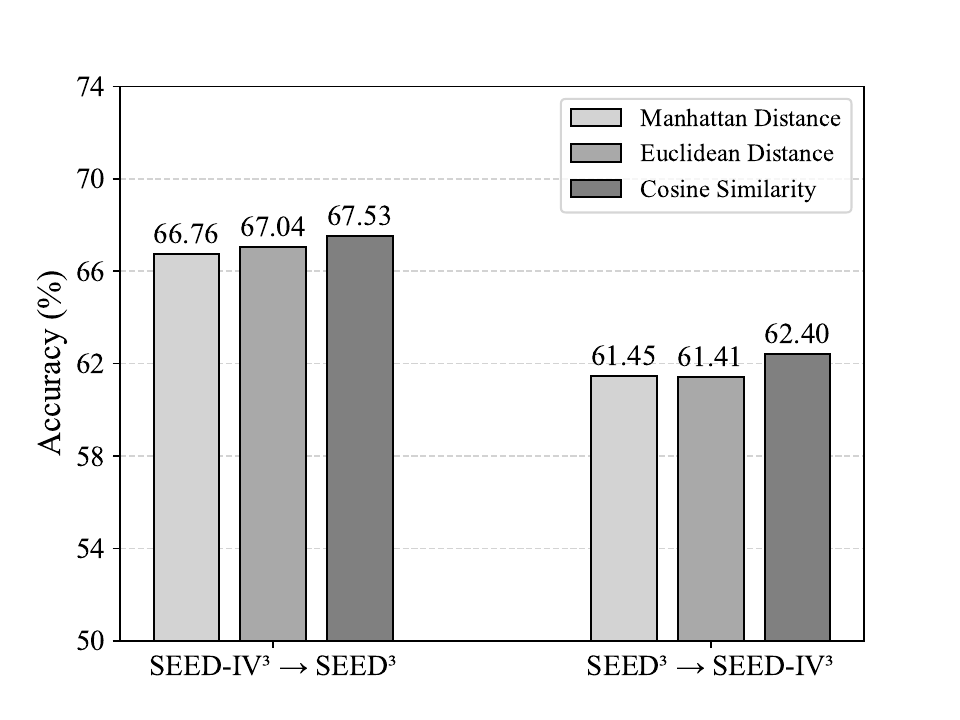}
    \end{center}
    \caption{Comparison of model performance on the SEED and SEED-IV datasets using different distance metrics.}
    \label{fig:Distance Metrics}
    \Description{This figure illustrates the model performance of the proposed JCFA for cross-corpus EEG-based emotion recognition using different distance metrics.}
\end{figure}

\subsection{Impact of Distance Metric}
We further assess the impact of the chosen distance metric in the construction of $G$ in the supervised fine-tuning stage. Three different distance metrics are individually employed in Eq. \ref{eq:Adjacency Matrix Based on Cosine Similarity}, and the corresponding results are shown in Fig. \ref{fig:Distance Metrics}. The comparison results indicate that the cosine similarity-based adjacency matrix is more suitable for constructing $G$ in the proposed JCFA model. This choice leads to the highest accuracies achieved in both SEED-IV$^3$ $\rightarrow$ SEED$^3$ and SEED$^3$ $\rightarrow$ SEED-IV$^3$ experiments.

\section{Conclusions}{
In this study, we propose a novel \textbf{J}oint \textbf{C}ontrastive learning framework with \textbf{F}eature \textbf{A}lignment (\textbf{JCFA}) to address the critical challenges associated with cross-corpus EEG-based emotion recognition. The JCFA model consists of two training stages. The pre-training stage is a self-supervised learning process aimed at efficiently capturing robust and generalizable time-frequency representations of raw EEG signals. A joint contrastive learning strategy is introduced to emphasize time- and frequency-based embeddings, all without depending on labeled data. Following the pre-training stage, the fine-tuning stage involves supervised learning with a graph convolutional network to enhance the model's capability for downstream tasks. Extensive experiments on two well-known datasets show that our approach achieves SOTA performance in comparison to existing methods. The model components and the adapted parameters are well examined, indicating that a consideration of time- and frequency-based embeddings alignment could be beneficial to EEG feature representation. The proposed JCFA model could be easily extended to other cross-corpus EEG tasks, further boosting and enhancing the practical applications of human-media interaction using brain signals.
} 

\bibliographystyle{ACM-Reference-Format}
\bibliography{sample-authordraft}


\begin{thebibliography}{46}

%%% the Web address.

\ifx \showCODEN    \undefined \def \showCODEN     #1{\unskip}     \fi
\ifx \showDOI      \undefined \def \showDOI       #1{#1}\fi
\ifx \showISBNx    \undefined \def \showISBNx     #1{\unskip}     \fi
\ifx \showISBNxiii \undefined \def \showISBNxiii  #1{\unskip}     \fi
\ifx \showISSN     \undefined \def \showISSN      #1{\unskip}     \fi
\ifx \showLCCN     \undefined \def \showLCCN      #1{\unskip}     \fi
\ifx \shownote     \undefined \def \shownote      #1{#1}          \fi
\ifx \showarticletitle \undefined \def \showarticletitle #1{#1}   \fi
\ifx \showURL      \undefined \def \showURL       {\relax}        \fi
% invisible to TeX
\providecommand\bibfield[2]{#2}
\providecommand\bibinfo[2]{#2}
\providecommand\natexlab[1]{#1}
\providecommand\showeprint[2][]{arXiv:#2}

\bibitem[Alarcao and Fonseca(2017)]%
        {alarcao2017emotions}
\bibfield{author}{\bibinfo{person}{Soraia~M Alarcao} {and} \bibinfo{person}{Manuel~J Fonseca}.} \bibinfo{year}{2017}\natexlab{}.
\newblock \showarticletitle{Emotions recognition using EEG signals: A survey}.
\newblock \bibinfo{journal}{\emph{IEEE transactions on affective computing}} \bibinfo{volume}{10}, \bibinfo{number}{3} (\bibinfo{year}{2017}), \bibinfo{pages}{374--393}.
\newblock


\bibitem[Alsolamy and Fattouh(2016)]%
        {alsolamy2016emotion}
\bibfield{author}{\bibinfo{person}{Mashail Alsolamy} {and} \bibinfo{person}{Anas Fattouh}.} \bibinfo{year}{2016}\natexlab{}.
\newblock \showarticletitle{Emotion estimation from EEG signals during listening to Quran using PSD features}. In \bibinfo{booktitle}{\emph{2016 7th International Conference on computer science and information technology (CSIT)}}. IEEE, \bibinfo{pages}{1--5}.
\newblock


\bibitem[Carpenter et~al\mbox{.}(2018)]%
        {carpenter2018cognitive}
\bibfield{author}{\bibinfo{person}{Joseph~K Carpenter}, \bibinfo{person}{Leigh~A Andrews}, \bibinfo{person}{Sara~M Witcraft}, \bibinfo{person}{Mark~B Powers}, \bibinfo{person}{Jasper~AJ Smits}, {and} \bibinfo{person}{Stefan~G Hofmann}.} \bibinfo{year}{2018}\natexlab{}.
\newblock \showarticletitle{Cognitive behavioral therapy for anxiety and related disorders: A meta-analysis of randomized placebo-controlled trials}.
\newblock \bibinfo{journal}{\emph{Depression and anxiety}} \bibinfo{volume}{35}, \bibinfo{number}{6} (\bibinfo{year}{2018}), \bibinfo{pages}{502--514}.
\newblock


\bibitem[Chen et~al\mbox{.}(2020)]%
        {chen2020simple}
\bibfield{author}{\bibinfo{person}{Ting Chen}, \bibinfo{person}{Simon Kornblith}, \bibinfo{person}{Mohammad Norouzi}, {and} \bibinfo{person}{Geoffrey Hinton}.} \bibinfo{year}{2020}\natexlab{}.
\newblock \showarticletitle{A simple framework for contrastive learning of visual representations}. In \bibinfo{booktitle}{\emph{International conference on machine learning}}. PMLR, \bibinfo{pages}{1597--1607}.
\newblock


\bibitem[Cowie et~al\mbox{.}(2001)]%
        {cowie2001emotion}
\bibfield{author}{\bibinfo{person}{Roddy Cowie}, \bibinfo{person}{Ellen Douglas-Cowie}, \bibinfo{person}{Nicolas Tsapatsoulis}, \bibinfo{person}{George Votsis}, \bibinfo{person}{Stefanos Kollias}, \bibinfo{person}{Winfried Fellenz}, {and} \bibinfo{person}{John~G Taylor}.} \bibinfo{year}{2001}\natexlab{}.
\newblock \showarticletitle{Emotion recognition in human-computer interaction}.
\newblock \bibinfo{journal}{\emph{IEEE Signal processing magazine}} \bibinfo{volume}{18}, \bibinfo{number}{1} (\bibinfo{year}{2001}), \bibinfo{pages}{32--80}.
\newblock


\bibitem[Defferrard et~al\mbox{.}(2016)]%
        {defferrard2016convolutional}
\bibfield{author}{\bibinfo{person}{Micha{\"e}l Defferrard}, \bibinfo{person}{Xavier Bresson}, {and} \bibinfo{person}{Pierre Vandergheynst}.} \bibinfo{year}{2016}\natexlab{}.
\newblock \showarticletitle{Convolutional neural networks on graphs with fast localized spectral filtering}.
\newblock \bibinfo{journal}{\emph{Advances in neural information processing systems}}  \bibinfo{volume}{29} (\bibinfo{year}{2016}).
\newblock


\bibitem[Duan et~al\mbox{.}(2013)]%
        {duan2013differential}
\bibfield{author}{\bibinfo{person}{Ruo-Nan Duan}, \bibinfo{person}{Jia-Yi Zhu}, {and} \bibinfo{person}{Bao-Liang Lu}.} \bibinfo{year}{2013}\natexlab{}.
\newblock \showarticletitle{Differential entropy feature for EEG-based emotion classification}. In \bibinfo{booktitle}{\emph{2013 6th international IEEE/EMBS conference on neural engineering (NER)}}. IEEE, \bibinfo{pages}{81--84}.
\newblock


\bibitem[Flandrin(1998)]%
        {flandrin1998time}
\bibfield{author}{\bibinfo{person}{Patrick Flandrin}.} \bibinfo{year}{1998}\natexlab{}.
\newblock \bibinfo{booktitle}{\emph{Time-frequency/time-scale analysis}}.
\newblock \bibinfo{publisher}{Academic press}.
\newblock


\bibitem[Fragopanagos and Taylor(2005)]%
        {fragopanagos2005emotion}
\bibfield{author}{\bibinfo{person}{Nickolaos Fragopanagos} {and} \bibinfo{person}{John~G Taylor}.} \bibinfo{year}{2005}\natexlab{}.
\newblock \showarticletitle{Emotion recognition in human--computer interaction}.
\newblock \bibinfo{journal}{\emph{Neural Networks}} \bibinfo{volume}{18}, \bibinfo{number}{4} (\bibinfo{year}{2005}), \bibinfo{pages}{389--405}.
\newblock


\bibitem[Ganin et~al\mbox{.}(2016)]%
        {ganin2016domain}
\bibfield{author}{\bibinfo{person}{Yaroslav Ganin}, \bibinfo{person}{Evgeniya Ustinova}, \bibinfo{person}{Hana Ajakan}, \bibinfo{person}{Pascal Germain}, \bibinfo{person}{Hugo Larochelle}, \bibinfo{person}{Fran{\c{c}}ois Laviolette}, \bibinfo{person}{Mario March}, {and} \bibinfo{person}{Victor Lempitsky}.} \bibinfo{year}{2016}\natexlab{}.
\newblock \showarticletitle{Domain-adversarial training of neural networks}.
\newblock \bibinfo{journal}{\emph{Journal of machine learning research}} \bibinfo{volume}{17}, \bibinfo{number}{59} (\bibinfo{year}{2016}), \bibinfo{pages}{1--35}.
\newblock


\bibitem[Goldfischer(1965)]%
        {goldfischer1965autocorrelation}
\bibfield{author}{\bibinfo{person}{Lester~I Goldfischer}.} \bibinfo{year}{1965}\natexlab{}.
\newblock \showarticletitle{Autocorrelation function and power spectral density of laser-produced speckle patterns}.
\newblock \bibinfo{journal}{\emph{Josa}} \bibinfo{volume}{55}, \bibinfo{number}{3} (\bibinfo{year}{1965}), \bibinfo{pages}{247--253}.
\newblock


\bibitem[Gong et~al\mbox{.}(2023)]%
        {gong2023astdf}
\bibfield{author}{\bibinfo{person}{Peiliang Gong}, \bibinfo{person}{Ziyu Jia}, \bibinfo{person}{Pengpai Wang}, \bibinfo{person}{Yueying Zhou}, {and} \bibinfo{person}{Daoqiang Zhang}.} \bibinfo{year}{2023}\natexlab{}.
\newblock \showarticletitle{ASTDF-Net: Attention-Based Spatial-Temporal Dual-Stream Fusion Network for EEG-Based Emotion Recognition}. In \bibinfo{booktitle}{\emph{Proceedings of the 31st ACM International Conference on Multimedia}}. \bibinfo{pages}{883--892}.
\newblock


\bibitem[Gu et~al\mbox{.}(2018)]%
        {gu2018deep}
\bibfield{author}{\bibinfo{person}{Yue Gu}, \bibinfo{person}{Shuhong Chen}, {and} \bibinfo{person}{Ivan Marsic}.} \bibinfo{year}{2018}\natexlab{}.
\newblock \showarticletitle{Deep mul timodal learning for emotion recognition in spoken language}. In \bibinfo{booktitle}{\emph{2018 IEEE International Conference on Acoustics, Speech and Signal Processing (ICASSP)}}. IEEE, \bibinfo{pages}{5079--5083}.
\newblock


\bibitem[Jin et~al\mbox{.}(2023)]%
        {jin2023pgcn}
\bibfield{author}{\bibinfo{person}{Ming Jin}, \bibinfo{person}{Enwei Zhu}, \bibinfo{person}{Changde Du}, \bibinfo{person}{Huiguang He}, {and} \bibinfo{person}{Jinpeng Li}.} \bibinfo{year}{2023}\natexlab{}.
\newblock \showarticletitle{PGCN: Pyramidal graph convolutional network for EEG emotion recognition}.
\newblock \bibinfo{journal}{\emph{arXiv preprint arXiv:2302.02520}} (\bibinfo{year}{2023}).
\newblock


\bibitem[Kiyasseh et~al\mbox{.}(2021)]%
        {kiyasseh2021clocs}
\bibfield{author}{\bibinfo{person}{Dani Kiyasseh}, \bibinfo{person}{Tingting Zhu}, {and} \bibinfo{person}{David~A Clifton}.} \bibinfo{year}{2021}\natexlab{}.
\newblock \showarticletitle{Clocs: Contrastive learning of cardiac signals across space, time, and patients}. In \bibinfo{booktitle}{\emph{International Conference on Machine Learning}}. PMLR, \bibinfo{pages}{5606--5615}.
\newblock


\bibitem[Lan et~al\mbox{.}(2018)]%
        {lan2018domain}
\bibfield{author}{\bibinfo{person}{Zirui Lan}, \bibinfo{person}{Olga Sourina}, \bibinfo{person}{Lipo Wang}, \bibinfo{person}{Reinhold Scherer}, {and} \bibinfo{person}{Gernot~R M{\"u}ller-Putz}.} \bibinfo{year}{2018}\natexlab{}.
\newblock \showarticletitle{Domain adaptation techniques for EEG-based emotion recognition: a comparative study on two public datasets}.
\newblock \bibinfo{journal}{\emph{IEEE Transactions on Cognitive and Developmental Systems}} \bibinfo{volume}{11}, \bibinfo{number}{1} (\bibinfo{year}{2018}), \bibinfo{pages}{85--94}.
\newblock


\bibitem[Li et~al\mbox{.}(2019)]%
        {li2019domain}
\bibfield{author}{\bibinfo{person}{Jinpeng Li}, \bibinfo{person}{Shuang Qiu}, \bibinfo{person}{Changde Du}, \bibinfo{person}{Yixin Wang}, {and} \bibinfo{person}{Huiguang He}.} \bibinfo{year}{2019}\natexlab{}.
\newblock \showarticletitle{Domain adaptation for EEG emotion recognition based on latent representation similarity}.
\newblock \bibinfo{journal}{\emph{IEEE Transactions on Cognitive and Developmental Systems}} \bibinfo{volume}{12}, \bibinfo{number}{2} (\bibinfo{year}{2019}), \bibinfo{pages}{344--353}.
\newblock


\bibitem[Li et~al\mbox{.}(2021b)]%
        {li2021multi}
\bibfield{author}{\bibinfo{person}{Rui Li}, \bibinfo{person}{Yiting Wang}, {and} \bibinfo{person}{Bao-Liang Lu}.} \bibinfo{year}{2021}\natexlab{b}.
\newblock \showarticletitle{A multi-domain adaptive graph convolutional network for EEG-based emotion recognition}. In \bibinfo{booktitle}{\emph{Proceedings of the 29th ACM International Conference on Multimedia}}. \bibinfo{pages}{5565--5573}.
\newblock


\bibitem[Li et~al\mbox{.}(2022)]%
        {li2022eeg}
\bibfield{author}{\bibinfo{person}{Xiang Li}, \bibinfo{person}{Yazhou Zhang}, \bibinfo{person}{Prayag Tiwari}, \bibinfo{person}{Dawei Song}, \bibinfo{person}{Bin Hu}, \bibinfo{person}{Meihong Yang}, \bibinfo{person}{Zhigang Zhao}, \bibinfo{person}{Neeraj Kumar}, {and} \bibinfo{person}{Pekka Marttinen}.} \bibinfo{year}{2022}\natexlab{}.
\newblock \showarticletitle{EEG based emotion recognition: A tutorial and review}.
\newblock \bibinfo{journal}{\emph{Comput. Surveys}} \bibinfo{volume}{55}, \bibinfo{number}{4} (\bibinfo{year}{2022}), \bibinfo{pages}{1--57}.
\newblock


\bibitem[Li et~al\mbox{.}(2021a)]%
        {li2021novel}
\bibfield{author}{\bibinfo{person}{Yang Li}, \bibinfo{person}{Boxun Fu}, \bibinfo{person}{Fu Li}, \bibinfo{person}{Guangming Shi}, {and} \bibinfo{person}{Wenming Zheng}.} \bibinfo{year}{2021}\natexlab{a}.
\newblock \showarticletitle{A novel transferability attention neural network model for EEG emotion recognition}.
\newblock \bibinfo{journal}{\emph{Neurocomputing}}  \bibinfo{volume}{447} (\bibinfo{year}{2021}), \bibinfo{pages}{92--101}.
\newblock


\bibitem[Li et~al\mbox{.}(2018)]%
        {li2018novel}
\bibfield{author}{\bibinfo{person}{Yang Li}, \bibinfo{person}{Wenming Zheng}, \bibinfo{person}{Zhen Cui}, \bibinfo{person}{Tong Zhang}, {and} \bibinfo{person}{Yuan Zong}.} \bibinfo{year}{2018}\natexlab{}.
\newblock \showarticletitle{A Novel Neural Network Model based on Cerebral Hemispheric Asymmetry for EEG Emotion Recognition}. In \bibinfo{booktitle}{\emph{IJCAI}}. \bibinfo{pages}{1561--1567}.
\newblock


\bibitem[Liu et~al\mbox{.}(2021)]%
        {liu2021contrastive}
\bibfield{author}{\bibinfo{person}{Dongxin Liu}, \bibinfo{person}{Tianshi Wang}, \bibinfo{person}{Shengzhong Liu}, \bibinfo{person}{Ruijie Wang}, \bibinfo{person}{Shuochao Yao}, {and} \bibinfo{person}{Tarek Abdelzaher}.} \bibinfo{year}{2021}\natexlab{}.
\newblock \showarticletitle{Contrastive self-supervised representation learning for sensing signals from the time-frequency perspective}. In \bibinfo{booktitle}{\emph{2021 International Conference on Computer Communications and Networks (ICCCN)}}. IEEE, \bibinfo{pages}{1--10}.
\newblock


\bibitem[Liu et~al\mbox{.}(2017)]%
        {liu2017real}
\bibfield{author}{\bibinfo{person}{Yong-Jin Liu}, \bibinfo{person}{Minjing Yu}, \bibinfo{person}{Guozhen Zhao}, \bibinfo{person}{Jinjing Song}, \bibinfo{person}{Yan Ge}, {and} \bibinfo{person}{Yuanchun Shi}.} \bibinfo{year}{2017}\natexlab{}.
\newblock \showarticletitle{Real-time movie-induced discrete emotion recognition from EEG signals}.
\newblock \bibinfo{journal}{\emph{IEEE Transactions on Affective Computing}} \bibinfo{volume}{9}, \bibinfo{number}{4} (\bibinfo{year}{2017}), \bibinfo{pages}{550--562}.
\newblock


\bibitem[Mohsenvand et~al\mbox{.}(2020)]%
        {mohsenvand2020contrastive}
\bibfield{author}{\bibinfo{person}{Mostafa~Neo Mohsenvand}, \bibinfo{person}{Mohammad~Rasool Izadi}, {and} \bibinfo{person}{Pattie Maes}.} \bibinfo{year}{2020}\natexlab{}.
\newblock \showarticletitle{Contrastive representation learning for electroencephalogram classification}. In \bibinfo{booktitle}{\emph{Machine Learning for Health}}. PMLR, \bibinfo{pages}{238--253}.
\newblock


\bibitem[Noroozi et~al\mbox{.}(2018)]%
        {noroozi2018survey}
\bibfield{author}{\bibinfo{person}{Fatemeh Noroozi}, \bibinfo{person}{Ciprian~Adrian Corneanu}, \bibinfo{person}{Dorota Kami{\'n}ska}, \bibinfo{person}{Tomasz Sapi{\'n}ski}, \bibinfo{person}{Sergio Escalera}, {and} \bibinfo{person}{Gholamreza Anbarjafari}.} \bibinfo{year}{2018}\natexlab{}.
\newblock \showarticletitle{Survey on emotional body gesture recognition}.
\newblock \bibinfo{journal}{\emph{IEEE transactions on affective computing}} \bibinfo{volume}{12}, \bibinfo{number}{2} (\bibinfo{year}{2018}), \bibinfo{pages}{505--523}.
\newblock


\bibitem[Nussbaumer and Nussbaumer(1982)]%
        {nussbaumer1982fast}
\bibfield{author}{\bibinfo{person}{Henri~J Nussbaumer} {and} \bibinfo{person}{Henri~J Nussbaumer}.} \bibinfo{year}{1982}\natexlab{}.
\newblock \bibinfo{booktitle}{\emph{The fast Fourier transform}}.
\newblock \bibinfo{publisher}{Springer}.
\newblock


\bibitem[Papandreou-Suppappola(2018)]%
        {papandreou2018applications}
\bibfield{author}{\bibinfo{person}{Antonia Papandreou-Suppappola}.} \bibinfo{year}{2018}\natexlab{}.
\newblock \bibinfo{booktitle}{\emph{Applications in time-frequency signal processing}}.
\newblock \bibinfo{publisher}{CRC press}.
\newblock


\bibitem[Rayatdoost and Soleymani(2018)]%
        {rayatdoost2018cross}
\bibfield{author}{\bibinfo{person}{Soheil Rayatdoost} {and} \bibinfo{person}{Mohammad Soleymani}.} \bibinfo{year}{2018}\natexlab{}.
\newblock \showarticletitle{Cross-corpus EEG-based emotion recognition}. In \bibinfo{booktitle}{\emph{2018 IEEE 28th international workshop on machine learning for signal processing (MLSP)}}. IEEE, \bibinfo{pages}{1--6}.
\newblock


\bibitem[Shen et~al\mbox{.}(2022)]%
        {shen2022contrastive}
\bibfield{author}{\bibinfo{person}{Xinke Shen}, \bibinfo{person}{Xianggen Liu}, \bibinfo{person}{Xin Hu}, \bibinfo{person}{Dan Zhang}, {and} \bibinfo{person}{Sen Song}.} \bibinfo{year}{2022}\natexlab{}.
\newblock \showarticletitle{Contrastive learning of subject-invariant EEG representations for cross-subject emotion recognition}.
\newblock \bibinfo{journal}{\emph{IEEE Transactions on Affective Computing}} (\bibinfo{year}{2022}).
\newblock


\bibitem[Song et~al\mbox{.}(2021a)]%
        {song2021variational}
\bibfield{author}{\bibinfo{person}{Tengfei Song}, \bibinfo{person}{Suyuan Liu}, \bibinfo{person}{Wenming Zheng}, \bibinfo{person}{Yuan Zong}, \bibinfo{person}{Zhen Cui}, \bibinfo{person}{Yang Li}, {and} \bibinfo{person}{Xiaoyan Zhou}.} \bibinfo{year}{2021}\natexlab{a}.
\newblock \showarticletitle{Variational instance-adaptive graph for EEG emotion recognition}.
\newblock \bibinfo{journal}{\emph{IEEE Transactions on Affective Computing}} \bibinfo{volume}{14}, \bibinfo{number}{1} (\bibinfo{year}{2021}), \bibinfo{pages}{343--356}.
\newblock


\bibitem[Song et~al\mbox{.}(2021b)]%
        {song2021graph}
\bibfield{author}{\bibinfo{person}{Tengfei Song}, \bibinfo{person}{Wenming Zheng}, \bibinfo{person}{Suyuan Liu}, \bibinfo{person}{Yuan Zong}, \bibinfo{person}{Zhen Cui}, {and} \bibinfo{person}{Yang Li}.} \bibinfo{year}{2021}\natexlab{b}.
\newblock \showarticletitle{Graph-embedded convolutional neural network for image-based EEG emotion recognition}.
\newblock \bibinfo{journal}{\emph{IEEE Transactions on Emerging Topics in Computing}} \bibinfo{volume}{10}, \bibinfo{number}{3} (\bibinfo{year}{2021}), \bibinfo{pages}{1399--1413}.
\newblock


\bibitem[Song et~al\mbox{.}(2019)]%
        {song2019mped}
\bibfield{author}{\bibinfo{person}{Tengfei Song}, \bibinfo{person}{Wenming Zheng}, \bibinfo{person}{Cheng Lu}, \bibinfo{person}{Yuan Zong}, \bibinfo{person}{Xilei Zhang}, {and} \bibinfo{person}{Zhen Cui}.} \bibinfo{year}{2019}\natexlab{}.
\newblock \showarticletitle{MPED: A multi-modal physiological emotion database for discrete emotion recognition}.
\newblock \bibinfo{journal}{\emph{IEEE Access}}  \bibinfo{volume}{7} (\bibinfo{year}{2019}), \bibinfo{pages}{12177--12191}.
\newblock


\bibitem[Song et~al\mbox{.}(2018)]%
        {song2018eeg}
\bibfield{author}{\bibinfo{person}{Tengfei Song}, \bibinfo{person}{Wenming Zheng}, \bibinfo{person}{Peng Song}, {and} \bibinfo{person}{Zhen Cui}.} \bibinfo{year}{2018}\natexlab{}.
\newblock \showarticletitle{EEG emotion recognition using dynamical graph convolutional neural networks}.
\newblock \bibinfo{journal}{\emph{IEEE Transactions on Affective Computing}} \bibinfo{volume}{11}, \bibinfo{number}{3} (\bibinfo{year}{2018}), \bibinfo{pages}{532--541}.
\newblock


\bibitem[Suykens and Vandewalle(1999)]%
        {suykens1999least}
\bibfield{author}{\bibinfo{person}{Johan~AK Suykens} {and} \bibinfo{person}{Joos Vandewalle}.} \bibinfo{year}{1999}\natexlab{}.
\newblock \showarticletitle{Least squares support vector machine classifiers}.
\newblock \bibinfo{journal}{\emph{Neural processing letters}}  \bibinfo{volume}{9} (\bibinfo{year}{1999}), \bibinfo{pages}{293--300}.
\newblock


\bibitem[Tang et~al\mbox{.}(2020)]%
        {tang2020exploring}
\bibfield{author}{\bibinfo{person}{Chi~Ian Tang}, \bibinfo{person}{Ignacio Perez-Pozuelo}, \bibinfo{person}{Dimitris Spathis}, {and} \bibinfo{person}{Cecilia Mascolo}.} \bibinfo{year}{2020}\natexlab{}.
\newblock \showarticletitle{Exploring Contrastive Learning in Human Activity Recognition for Healthcare}.
\newblock \bibinfo{journal}{\emph{arXiv preprint arXiv:2011.11542}} (\bibinfo{year}{2020}).
\newblock


\bibitem[Van~der Maaten and Hinton(2008)]%
        {van2008visualizing}
\bibfield{author}{\bibinfo{person}{Laurens Van~der Maaten} {and} \bibinfo{person}{Geoffrey Hinton}.} \bibinfo{year}{2008}\natexlab{}.
\newblock \showarticletitle{Visualizing data using t-SNE.}
\newblock \bibinfo{journal}{\emph{Journal of machine learning research}} \bibinfo{volume}{9}, \bibinfo{number}{11} (\bibinfo{year}{2008}).
\newblock


\bibitem[Wickstr{\o}m et~al\mbox{.}(2022)]%
        {wickstrom2022mixing}
\bibfield{author}{\bibinfo{person}{Kristoffer Wickstr{\o}m}, \bibinfo{person}{Michael Kampffmeyer}, \bibinfo{person}{Karl~{\O}yvind Mikalsen}, {and} \bibinfo{person}{Robert Jenssen}.} \bibinfo{year}{2022}\natexlab{}.
\newblock \showarticletitle{Mixing up contrastive learning: Self-supervised representation learning for time series}.
\newblock \bibinfo{journal}{\emph{Pattern Recognition Letters}}  \bibinfo{volume}{155} (\bibinfo{year}{2022}), \bibinfo{pages}{54--61}.
\newblock


\bibitem[Yue et~al\mbox{.}(2022)]%
        {yue2022ts2vec}
\bibfield{author}{\bibinfo{person}{Zhihan Yue}, \bibinfo{person}{Yujing Wang}, \bibinfo{person}{Juanyong Duan}, \bibinfo{person}{Tianmeng Yang}, \bibinfo{person}{Congrui Huang}, \bibinfo{person}{Yunhai Tong}, {and} \bibinfo{person}{Bixiong Xu}.} \bibinfo{year}{2022}\natexlab{}.
\newblock \showarticletitle{Ts2vec: Towards universal representation of time series}. In \bibinfo{booktitle}{\emph{Proceedings of the AAAI Conference on Artificial Intelligence}}, Vol.~\bibinfo{volume}{36}. \bibinfo{pages}{8980--8987}.
\newblock


\bibitem[Zeng et~al\mbox{.}(2018)]%
        {zeng2018facial}
\bibfield{author}{\bibinfo{person}{Nianyin Zeng}, \bibinfo{person}{Hong Zhang}, \bibinfo{person}{Baoye Song}, \bibinfo{person}{Weibo Liu}, \bibinfo{person}{Yurong Li}, {and} \bibinfo{person}{Abdullah~M Dobaie}.} \bibinfo{year}{2018}\natexlab{}.
\newblock \showarticletitle{Facial expression recognition via learning deep sparse autoencoders}.
\newblock \bibinfo{journal}{\emph{Neurocomputing}}  \bibinfo{volume}{273} (\bibinfo{year}{2018}), \bibinfo{pages}{643--649}.
\newblock


\bibitem[Zhang et~al\mbox{.}(2017)]%
        {zhang2017mixup}
\bibfield{author}{\bibinfo{person}{Hongyi Zhang}, \bibinfo{person}{Moustapha Cisse}, \bibinfo{person}{Yann~N Dauphin}, {and} \bibinfo{person}{David Lopez-Paz}.} \bibinfo{year}{2017}\natexlab{}.
\newblock \showarticletitle{mixup: Beyond empirical risk minimization}.
\newblock \bibinfo{journal}{\emph{arXiv preprint arXiv:1710.09412}} (\bibinfo{year}{2017}).
\newblock


\bibitem[Zheng et~al\mbox{.}(2018)]%
        {zheng2018emotionmeter}
\bibfield{author}{\bibinfo{person}{Wei-Long Zheng}, \bibinfo{person}{Wei Liu}, \bibinfo{person}{Yifei Lu}, \bibinfo{person}{Bao-Liang Lu}, {and} \bibinfo{person}{Andrzej Cichocki}.} \bibinfo{year}{2018}\natexlab{}.
\newblock \showarticletitle{Emotionmeter: A multimodal framework for recognizing human emotions}.
\newblock \bibinfo{journal}{\emph{IEEE transactions on cybernetics}} \bibinfo{volume}{49}, \bibinfo{number}{3} (\bibinfo{year}{2018}), \bibinfo{pages}{1110--1122}.
\newblock


\bibitem[Zheng and Lu(2015)]%
        {zheng2015investigating}
\bibfield{author}{\bibinfo{person}{Wei-Long Zheng} {and} \bibinfo{person}{Bao-Liang Lu}.} \bibinfo{year}{2015}\natexlab{}.
\newblock \showarticletitle{Investigating critical frequency bands and channels for EEG-based emotion recognition with deep neural networks}.
\newblock \bibinfo{journal}{\emph{IEEE Transactions on autonomous mental development}} \bibinfo{volume}{7}, \bibinfo{number}{3} (\bibinfo{year}{2015}), \bibinfo{pages}{162--175}.
\newblock


\bibitem[Zhong et~al\mbox{.}(2020)]%
        {zhong2020eeg}
\bibfield{author}{\bibinfo{person}{Peixiang Zhong}, \bibinfo{person}{Di Wang}, {and} \bibinfo{person}{Chunyan Miao}.} \bibinfo{year}{2020}\natexlab{}.
\newblock \showarticletitle{EEG-based emotion recognition using regularized graph neural networks}.
\newblock \bibinfo{journal}{\emph{IEEE Transactions on Affective Computing}} \bibinfo{volume}{13}, \bibinfo{number}{3} (\bibinfo{year}{2020}), \bibinfo{pages}{1290--1301}.
\newblock


\bibitem[Zhou et~al\mbox{.}(2023b)]%
        {zhou2023pr}
\bibfield{author}{\bibinfo{person}{Rushuang Zhou}, \bibinfo{person}{Zhiguo Zhang}, \bibinfo{person}{Hong Fu}, \bibinfo{person}{Li Zhang}, \bibinfo{person}{Linling Li}, \bibinfo{person}{Gan Huang}, \bibinfo{person}{Fali Li}, \bibinfo{person}{Xin Yang}, \bibinfo{person}{Yining Dong}, \bibinfo{person}{Yuan-Ting Zhang}, {et~al\mbox{.}}} \bibinfo{year}{2023}\natexlab{b}.
\newblock \showarticletitle{PR-PL: A novel prototypical representation based pairwise learning framework for emotion recognition using EEG signals}.
\newblock \bibinfo{journal}{\emph{IEEE Transactions on Affective Computing}} (\bibinfo{year}{2023}).
\newblock


\bibitem[Zhou et~al\mbox{.}(2023a)]%
        {zhou2023eeg}
\bibfield{author}{\bibinfo{person}{Yijin Zhou}, \bibinfo{person}{Fu Li}, \bibinfo{person}{Yang Li}, \bibinfo{person}{Youshuo Ji}, \bibinfo{person}{Lijian Zhang}, \bibinfo{person}{Yuanfang Chen}, \bibinfo{person}{Wenming Zheng}, {and} \bibinfo{person}{Guangming Shi}.} \bibinfo{year}{2023}\natexlab{a}.
\newblock \showarticletitle{EEG-based Emotion Style Transfer Network for Cross-dataset Emotion Recognition}.
\newblock \bibinfo{journal}{\emph{arXiv preprint arXiv:2308.05767}} (\bibinfo{year}{2023}).
\newblock


\bibitem[Zotev et~al\mbox{.}(2020)]%
        {zotev2020emotion}
\bibfield{author}{\bibinfo{person}{Vadim Zotev}, \bibinfo{person}{Ahmad Mayeli}, \bibinfo{person}{Masaya Misaki}, {and} \bibinfo{person}{Jerzy Bodurka}.} \bibinfo{year}{2020}\natexlab{}.
\newblock \showarticletitle{Emotion self-regulation training in major depressive disorder using simultaneous real-time fMRI and EEG neurofeedback}.
\newblock \bibinfo{journal}{\emph{NeuroImage: Clinical}}  \bibinfo{volume}{27} (\bibinfo{year}{2020}), \bibinfo{pages}{102331}.
\newblock


\end{thebibliography}

\appendix
\renewcommand{\thesection}{Appendix \Alph{section}}
\renewcommand{\thetable}{A-\arabic{table}}
\renewcommand{\thefigure}{A-\arabic{figure}}
\setcounter{table}{0}
\setcounter{figure}{0}

\section{Datasets}
\label{appendix:Datasets}
{The SEED \cite{zheng2015investigating} and SEED-IV \cite{zheng2018emotionmeter} datasets are developed by the BCMI laboratory at SJTU. Both datasets used the 62-channel ESI NeuroScan System2 based on the international 10-20 system to record EEG signals of subjects under different types of video stimuli. The raw EEG signals were collected at a sampling rate of 1000Hz for the SEED and SEED-IV datasets. Table \ref{tab:Data statistics} shows the data statistics of the two datasets.

Specifically, the \textbf{SEED dataset} records EEG signals of 15 subjects (7 males and 8 females) under different video stimuli. Each subject participated in three different sessions. In each session, each subject was required to watch 15 movie clips containing 3 different emotional states (i.e., negative, neutral and positive emotions). Each emotion contains a total of 5 movie clips. The \textbf{SEED-IV dataset} is similar to the SEED dataset, which records EEG signals of 15 subjects participating in three different sessions. Each session has 24 trials corresponding to 24 movie clips which are evenly divided into 4 emotional states (i.e., sad, neutral, fear and happy emotions). In the experiments, we only use the preprocessed 1-s EEG signals from session 1 for both datasets.

The raw EEG signals from the SEED and SEED-IV datasets are first preprocessed through a common process. Specifically, the raw data is first downsampled to a sampling rate of 200Hz, and then filtered through a 1-75Hz bandpass filter to remove the noise and artifacts. After that, we divide the preprocessed EEG signals into multiple segments using a sliding window with a length of 1s. Therefore, the timestamps $T^{\mathrm{pret}}$ and $T^{\mathrm{tune}}$ of the input EEG samples are both 200. In the experiments, we only consider the case where the pre-training and fine-tuning datasets have the same emotion categories. However, the SEED-IV dataset contains four emotional states as shown in Table \ref{tab:Data statistics}. As a result, we exclude all samples corresponding to the fear emotions in the SEED-IV dataset. Finally, the SEED and SEED-IV experimental datasets only contain preprocessed 1-s EEG samples from 15 subjects in the first session associated with three emotion categories (i.e., negative, neutral and positive emotions). In particular, the SEED dataset contains 50910 samples, where each subject has 3394 samples corresponding to 15 trials. The SEED-IV dataset has a total of 40440 samples, where each subject contains 2696 samples corresponding to 18 trials. Notably, our approach is equally applicable to the case where the number of emotion categories is different in the pre-training and fine-tuning datasets, but we do not consider it in this paper.
\begin{table*}
\caption{Detailed description of the SEED and SEED-IV datasets.}
\centering
\renewcommand{\arraystretch}{}
    \begin{threeparttable}
    \begin{tabular}{c|c|c|c|c|c|c}
    \hline
    \hline
        Datasets & Subjects & Sessions & Trials & Channels & Sampling Rate (Hz) & Classes \\
        \hline
        SEED & 15 & 3 & 15 (3 $\times$ 5) & 62 & 1000 & 3 (negative, neutral, positive)\\
        SEED-IV & 15 & 3 & 24 (4 $\times$ 6) & 62 & 1000 & 4 (sad, neutral, fear, happy)\\ 
    \hline
    \hline
    \end{tabular}
    \begin{tablenotes}
        \footnotesize 
        \item A $\times$ B indicates that the number of classes is A, and each class contains B trials.
    \end{tablenotes}
    \end{threeparttable}
\label{tab:Data statistics}
\end{table*}
}

\section{Implementation Details}
\label{appendix:Implementation details}
{\subsection{Pre-training} 
\label{subsec:Pre-training} The time-based contrastive encoder $E_{\rm T}$ adopts a 2-layer transformer encoder, while the cross-space projector $P_{\rm T}$ uses a 2-layer fully connected network. The output feature dimensions for $E_{\rm T}$ and $P_{\rm T}$ are 200 and 128, respectively. The frequency-based contrastive encoder $E_{\rm F}$ and projector $P_{\rm F}$ have the same structures as $E_{\rm T}$ and $P_{\rm T}$, but with different parameters. Due to the non-stationarity of raw EEG signals, we set $\tau$ to the same and small value of 0.05 throughout all experiments to increase the penalization of hard negative samples. Then, we set $\alpha$ to 0.2 and $\beta$ to 1. We use Adam as the optimizer with learning rate of 3e-4 and L2-norm penalty coefficient of 3e-4. The number of pre-training epochs is set to 1000 and the batch size is 256. We save the model parameters corresponding to the last epoch as the pre-trained model $\mathcal{F}$.

\subsection{Fine-tuning}
\label{subsec:Fine-tuning} For the Chebyshev graph convolutional network $G$, we set $\delta$ to a small value of 0.2 to avoid introducing excessive useless information. Then, we set the node feature dimensions of the first output layer to 128 and the second layer to 64. We set $K$ to a small value of 3 to avoid over-smoothing. For the emotion classifier $C$, we adopt a 3-layer fully connected network with hidden dimensions 1024, 128 and the number of classes in the fine-tuning dataset, respectively. Differing from the pre-training process, we set $\tau$ to a relative large value of 0.5, and we set both $\alpha$ and $\beta$ to 0.1 and $\gamma$ to 1. We use Adam optimizer with the initial learning rate of 5e-4 and weight decay of 3e-4. We set batch size to 128 and fine-tuning epochs to 20. During the model fine-tuning process, we record the model performance for each epoch and save the corresponding parameters according to the best classification accuracy. Finally, the fine-tuned model is used for emotion recognition on all test samples.

\section{Baseline Models}
\label{appendix:Baselines}
We compare the JCFA model with 8 conventional methods and 2 classical contrastive learning models. Throughout the experiments, we use the default hyper-parameters reported in the original paper, and follow the same experimental settings for each method to ensure a fair comparison. Details of the 10 baseline methods are described below:
            
\begin{itemize}
\item SVM \cite{suykens1999least}: A support vector machine with the linear kernel. It has been fully tested on the SEED and SEED-IV datasets.
\item DANN \cite{ganin2016domain}: The domain adversarial neural network is a typical model in the field of transfer learning, which has been extended to several EEG emotion datasets to demonstrate its effectiveness in data migration.
\item A-LSTM \cite{song2019mped}: A novel approach for extracting discriminative features using attention-long short-term memory from multiple time-frequency domain features.
\item V-IAG \cite{song2021variational}: The variational-instance adaptive graph method simultaneously captures the individual dependencies among different EEG electrodes and estimates the underlying uncertain information.
\item GECNN \cite{song2021graph}: A novel graph-embedded convolutional neural network that converts the question of EEG-based emotion recognition into image recognition.
\item BiDANN \cite{li2018novel}: The bi-hemispheres domain adversarial neural network is developed to address domain shifts in cross-subject EEG-based emotion recognition tasks.
\item TANN \cite{li2021novel}: The transferable attention neural network is a novel transfer learning method that learns the discriminative information from EEG signals based on the local and global attention mechanism.
\item E$^2$STN \cite{zhou2023eeg}: The EEG-based emotion style transfer network integrates content information from the source domain with style information from the target domain, and achieves state-of-the-art performance in multiple cross-corpus scenarios.
\item SimCLR \cite{chen2020simple, tang2020exploring}: A groundbreaking work in self-supervised representation learning of images, which has been extended to human activity recognition in \cite{tang2020exploring}. 
\item Mixup \cite{zhang2017mixup, wickstrom2022mixing}: A classical contrastive learning model which proposes a novel data augmentation method, and it has been extended to time series analysis in \cite{wickstrom2022mixing}.
\end{itemize}

During the experiments, we adopt the cross-corpus subject-independent protocol. Specifically, for the first 8 methods, we use the DE features extracted from the raw EEG signals as input. The samples of all subjects from one dataset are used as source training set, and the samples of each subject from another dataset are considered separately as target test set. For SimCLR and Mixup, we first modify their model structure to a two-stage model by adding an emotion classifier consisting of a 3-layer fully connected network in the fine-tuning stage. Then, we use the preprocessed 1-s EEG signals as input, and divide datasets into source training set and target set. The target set is further divided into fine-tuning set and test set. After that, we use the leave-trials-out setting in the fine-tuning process, that is, the samples from a part of the trials of each subject in the target set are used for model fine-tuning, and the remaining trials are used for testing. In this way, we can effectively avoid data leakage.
}

\section{Visualization of the Learned Node Features in $G$}
\label{appendix:visualization}
\begin{figure}[h]
    \begin{center}
        \includegraphics[width=0.25\textwidth]{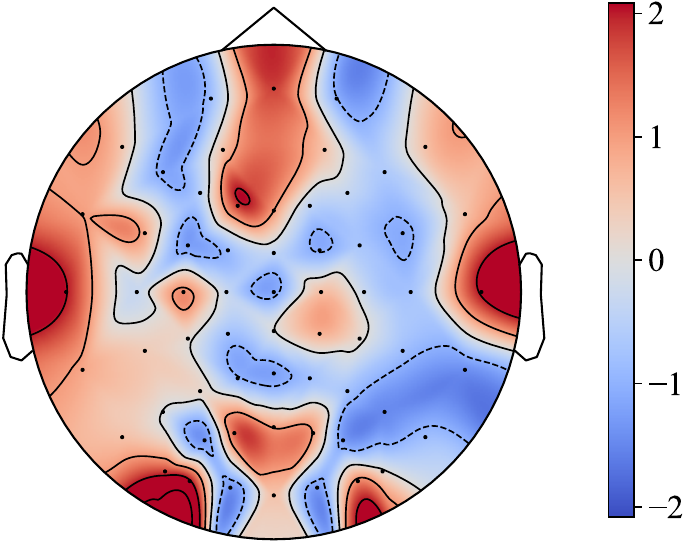}
    \end{center}
    \caption{Heat map of the learned node features in $G$.}
    \Description{Heat map of the learned node features in $G$.}
    \label{fig:topomap}
\end{figure}

To further explore the contribution of different brain regions for EEG-based emotion recognition, we select one subject from the SEED dataset and use the heat map to visualize the learned node features in graph $G$. Figure \ref{fig:topomap} depicts the result of the node features visualization. The darker red areas indicate that brain electrodes contribute more in the corresponding regions. We can clearly observe that brain electrodes in the frontal and temporal lobes gain higher weights and contribute more for cross-corpus EEG-based emotion recognition, which is consistent with existing research in neuroscience \cite{alarcao2017emotions}. This demonstrates the effectiveness of the Chebyshev graph convolutional network $G$ we designed in the fine-tuning stage, which is able to extract the spatial features related to the emotion categories of brain electrodes, thereby improving the classification performance for different emotion categories.

\section{Loss Function Curves of Model Training}
\begin{figure}[h]
    \begin{center}
        \subfloat{
            \begin{minipage}[c]{0.22\textwidth}
            \centering
                \includegraphics[width=1\textwidth]{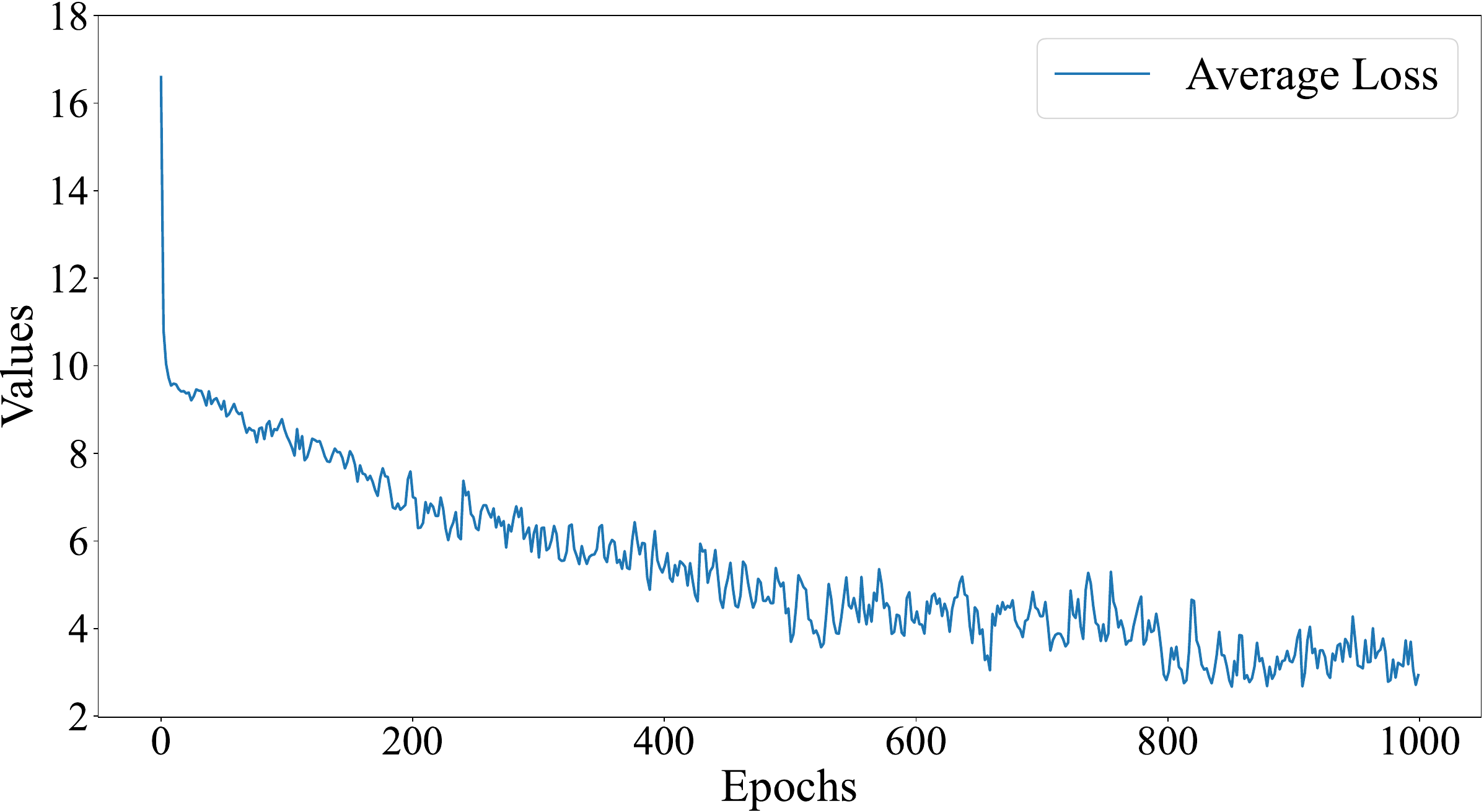}
                (a)
            \end{minipage}
            }
        \hspace{0.5em}
        \subfloat{
            \begin{minipage}[c]{0.22\textwidth}
            \centering
                \includegraphics[width=1\textwidth]{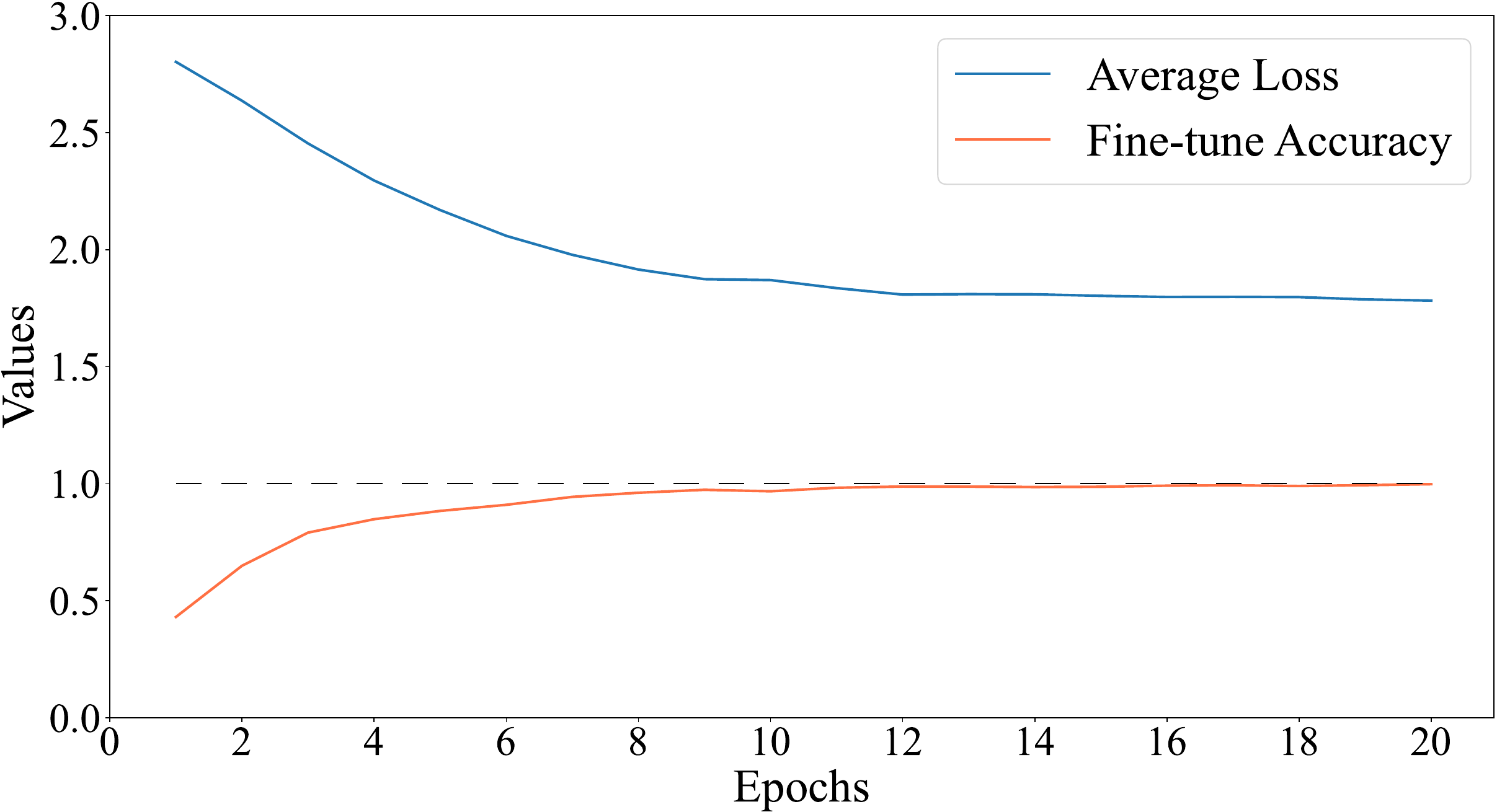}
                (b)
            \end{minipage}
            }
        \caption{Loss function curves of model training. (a) is the pre-training stage and (b) is the fine-tuning stage.}
    \label{fig:loss curves}
    \end{center}
    \Description{Loss function curves of model training.}
\end{figure}

Figure \ref{fig:loss curves} depicts the loss function curves for our proposed JCFA model in the pre-training and fine-tuning stages. 

\end{document}